# Surface Effect on Domain Wall Width in Ferroelectrics

Eugene A. Eliseev,[1, *, ‡] Anna N. Morozovska,[1, †, ‡] Sergei V. Kalinin,[2] Yulan L. Li,[3] Jie Shen,[4] Maya D. Glinchuk,[1] Long-Qing Chen,[3] and Venkatraman Gopalan[3]

[1]Institute for Problems of Materials Science, National Academy of Science of Ukraine, 3, Krjijanovskogo, 03142 Kiev, Ukraine

[2]The Center for Nanophase Materials Sciences and Materials Science and Technology Division, Oak Ridge National Laboratory, Oak Ridge Tennessee 37831, USA

[3] Department of Materials Science and Engineering, Pennsylvania State University, University Park, Pennsylvania 16802, USA

[4]Department of Mathematics, Purdue University, West Lafayette, Indiana 47907, USA

**Abstract**

We study the effect of depolarization field related with inhomogeneous polarization distribution, strain and surface energy parameters on a domain wall profile near the surface of a ferroelectric film within the framework of Landau-Ginzburg-Devonshire phenomenology. Both inhomogeneous elastic stress and positive surface energy lead to the wall broadening at electrically screened surface. For ferroelectrics with weak piezoelectric coupling, the extrapolation length that defines surface energy parameter, affects the wall broadening more strongly than inhomogeneous elastic stress. Unexpectedly, the domain wall profile follows a long-range power law when approaching the surface, while it saturates exponentially in the bulk. In materials with high piezoelectric coupling and negligibly small surface energy (i.e. high extrapolation length) inhomogeneous elastic stress effect dominates.

[*] E-mail: eliseev@i.com.ua

[†] E-mail: morozo@i.com.ua, permanent address: V.Lashkaryov Institute of Semiconductor Physics, National Academy of Science of Ukraine, 41, pr. Nauki, 03028 Kiev, Ukraine

[‡] These authors contributed equally to this work.

## 1. Introduction

Surfaces and interfaces in ferroic materials have been attracting much attention since early seventies till the present.[1, 2, 3, 4, 5, 6, 7] Laminar domain structure formation in the thick films with free surfaces was considered in the classic papers by Kittel[8] for ferromagnetic media and Mitsui and Furuichi[9] for ferroelectric media. In the classical papers of Cao and Cross[10] and Zhirnov[11], a single boundary between two domains was considered in the bulk ferroelectrics, allowing for electrostriction contribution. The stability of domain structure in the electroded thin films in the presence of dielectric dead layer was considered by Bratkovsky and Levanyuk,[12] while Pertsev and Kohlstedt[13] considered finite screening length in electrodes.

The development of non-volatile ferroelectric memory technology has rekindled the interest in ferroelectric properties and polarization reversal mechanisms in ultrathin films.[14, 15, 16, 17] One of the key parameters controlling ferroic behavior in thin films is the structure and energetic of domain walls. Wall energy determines the relative stability of mono- vs. periodic domain states, thus directly controlling phase-field-temperature behavior of the material. Furthermore, the wall behavior and energetics at surfaces and interfaces will determine polarization switching and pinning mechanisms. Under the absence of external fields in the bulk, the 180°-domain wall is not associated with any depolarization effects. However, the symmetry breaking on the wall-surface or wall-interface junction can give rise to a variety of unusual effects due to the depolarization fields across the wall, as determined by screening mechanism and strain boundary conditions. Recent Density Functional Theory (DFT) simulation results predicted the stabilization of vortex structure in ferroelectric nanodots under the transverse inhomogeneous static electric field;[18] and the existence of the tilted closure domains and bubble domain states on the free surfaces.[19] Prosandeev and Bellaiche[19] considered formation of various domain structures under the asymmetric electric boundary conditions. Recent results in this field are presented in Ref.[20] The in-plane polarization component at the wall-surface junction lowers the associated electrostatic depolarization energy. This prediction has resulted in extensive experimental effort to discover toroidal polarization states in ferroelectrics.[21, 22]

Formation and stability of 90° domain structures in epitaxial films of ferroelectric – insipient ferroelastic under the misfit strain influence were considered by different groups, with special attention to the period and conditions of existence [23, 24, 25] and transition between monodomain and polydomain structures.[26] However, the details of domain walls structure were not considered in these studies. Domain walls structure in proper ferroelastic was considered by Salje et al.[27, 28, 29] It was found[28] that nonzero internal stress exists at a domain wall due to coupling between primary order parameter (shear strain) and dilatation strains there. Relaxation of stress normal components at the free surface led to either the wall widening or narrowing near the surface depending on the sign of surface curvature. The role of point defects on the domain wall widths was analyzed by Lee et al.[29]

Despite the enormous progress achieved in atomistic and DFT modeling of multidomain ferroelectrics and domain wall behavior in thin films and its significant relevance to virtually all aspects of ferroelectric and other ferroic materials, the question of near-surface structure of a domain boundary in the proper ferroelectrics was virtually out of consideration, the only exception is the early work of Darinskii et al.[30] It was found that near the surface the ratio of the saturation value to the slope in the wall center is exactly the same as in the bulk. However, they did not consider the changes of domain wall profile near the surface.

The intrinsic 2D nature of the problem (as compared to 1D capacitors) and non-linear equations of state inherent in GLD description rendered this problem poorly amenable to direct analytical treatment. Consequently, polarization and/or elastic strain in the multidomain state are typically considered using harmonic function approximation,[2, 13, 26] which is indeed reasonable near the phase transition point, but this approximation could not allow one to grasp the domain walls details for the case of developed domain structure. In contrast, in the paper we derive the domain wall profile near the surface self-consistently using perturbation theory.

Paper is organized as following. In Sections 2 and 3 we present general approach for depolarization field calculations and Euler-Lagrange equation. Domain wall surface broadening caused by inhomogeneous elastic stress is considered in Section 4. Domain wall broadening caused by finite extrapolation length and comparison with available experimental data is considered in Section 5.

## 2. Depolarization field

The Maxwell's equation $\operatorname{div} \mathbf{D} = 0$ for the displacement $\mathbf{D} = \mathbf{P}(\mathbf{r}) - \varepsilon_0 \nabla \varphi(\mathbf{r})$ expressed via electrostatic potential $\varphi(\mathbf{r})$ and polarization $\mathbf{P}(\mathbf{r})$ with short-circuited boundary conditions have the form:

$$\begin{cases} \operatorname{div}(\mathbf{P}(\mathbf{r}) - \varepsilon_0 \nabla \varphi(\mathbf{r})) = 0, & z \geq 0, \\ \varphi(x, y, 0) = 0, & \varphi(x, y, h) = 0 \end{cases} \quad (1)$$

Electrostatic potential $\varphi(\mathbf{r})$ includes bond charges (electric depolarization field); $\varepsilon_0$ is the dielectric constant, $h$ is the film thickness [see Fig.1]. The perfect screening of depolarization field outside the sample is assumed by the ambient charges or free charges in electrodes. The effect of incomplete screening due to the presence of dead layer[2, 10, 31], and finite screening length of electrodes[5, 13] on the domain wall-surface junction is not considered in the present work. Rather, we concentrate on the ideal screened case.

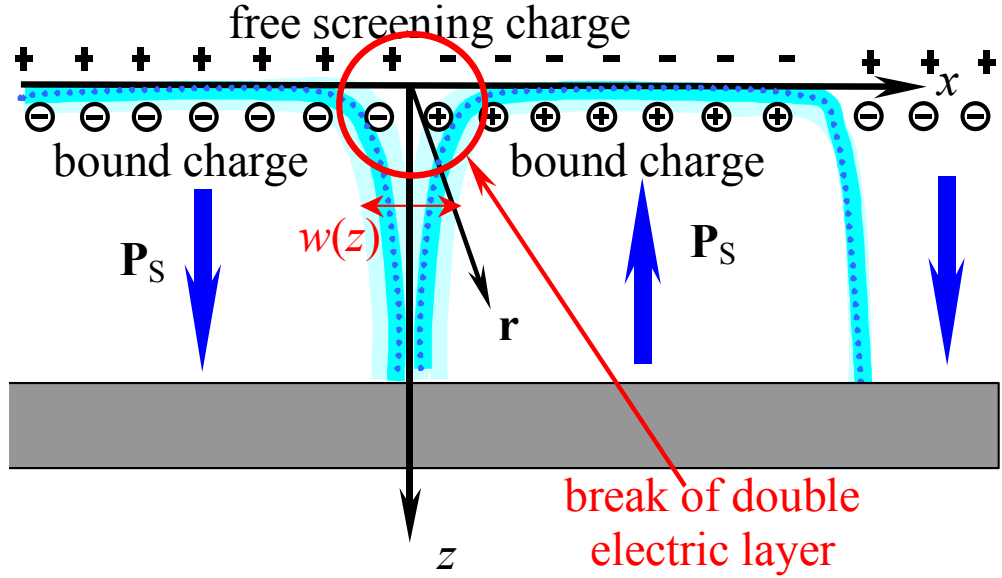

**Figure 1.** (Color online) 180º-domain structure near the film surface. Break of double electric layer causes stray depolarization field.

Hereinafter we consider uniaxial ferroelectrics with initial spontaneous polarization $P_3^S$ directed along the polar axis z. The sample is dielectrically isotropic in transverse directions, i.e.

permittivity $\varepsilon_{11} = \varepsilon_{22}$ at zero external field. We assume that the dependence of in-plane polarization components on the inner field $E_{1,2}$ can be linearized as $P_{1,2} \approx \varepsilon_0(\varepsilon_{11} - 1)E_{1,2}$. Thus the polarization vector acquires the form: $\mathbf{P}(\mathbf{r}) = (\varepsilon_0(\varepsilon_{11} - 1)E_1,\ \varepsilon_0(\varepsilon_{11} - 1)E_2,\ P_3(\mathbf{E},\mathbf{r}))$.[32]

We can rewrite the problem (1) for electrostatic potential as:

$$\begin{cases} \dfrac{\partial^2 \varphi}{\partial z^2} + \varepsilon_{11}\left(\dfrac{\partial^2 \varphi}{\partial x^2} + \dfrac{\partial^2 \varphi}{\partial y^2}\right) = \dfrac{1}{\varepsilon_0}\dfrac{\partial P_3}{\partial z} \\ \varphi(x,y,0) = 0, \quad \varphi(x,y,h) = 0 \end{cases} \quad (2)$$

Corresponding Fourier representation on transverse coordinates $\{x,y\}$ for electrostatic potential $\widetilde{\varphi}(\mathbf{k}, z)$ and electric field normal component $\widetilde{E}_3 = -\partial \widetilde{\varphi}/\partial z$ have the form:

$$\widetilde{\varphi}(\mathbf{k},z) = \begin{pmatrix} \int_0^z d\xi \widetilde{P}_3(\mathbf{k},\xi) \dfrac{\cosh(k\xi\sqrt{\varepsilon_{11}})\sinh(k(h-z)\sqrt{\varepsilon_{11}})}{\varepsilon_0 \cdot \sinh(k h\sqrt{\varepsilon_{11}})} - \\ \int_z^h d\xi \widetilde{P}_3(\mathbf{k},\xi) \dfrac{\sinh(k z\sqrt{\varepsilon_{11}})\cosh(k(h-\xi)\sqrt{\varepsilon_{11}})}{\varepsilon_0 \cdot \sinh(k h\sqrt{\varepsilon_{11}})} \end{pmatrix} \quad (3)$$

$$\widetilde{E}_3^d(\mathbf{k},z) = \begin{pmatrix} -\dfrac{1}{\varepsilon_0}\widetilde{P}_3(\mathbf{k},z) + \\ +\int_0^z d\xi \widetilde{P}_3(\mathbf{k},\xi)\cosh(k\xi\sqrt{\varepsilon_{11}})\dfrac{\cosh(k(h-z)\sqrt{\varepsilon_{11}})}{\varepsilon_0 \cdot \sinh(k h\sqrt{\varepsilon_{11}})}k\sqrt{\varepsilon_{11}} + \\ \int_z^h d\xi \widetilde{P}_3(\mathbf{k},\xi)\cosh(k(h-\xi)\sqrt{\varepsilon_{11}})\dfrac{\cosh(k z\sqrt{\varepsilon_{11}})k\sqrt{\varepsilon_{11}}}{\varepsilon_0 \cdot \sinh(k h\sqrt{\varepsilon_{11}})} \end{pmatrix} \quad (4)$$

Here vector $\mathbf{k} = \{k_1, k_2\}$, its absolute value $k = \sqrt{k_1^2 + k_2^2}$. Ever under the perfect external screening or short-circuit condition the field (4) is nonzero; it is produced by inhomogeneous polarization distribution, i.e. it is a typical depolarization field $E_3^d$ denoted by superscript "$d$". Note that for the transversally homogeneous media Eq.(4) reduces to the expression for depolarization field obtained by Kretschmer and Binder[1] (see Appendix A for details).

## 3. Euler-Lagrange equation

Spatial distribution of the polarization component $P_3$ inside the sample could be found by direct variational method[33, 34] of the free energy functional minimization:

$$G = \int_0^h dz \int_{-\infty}^{\infty} dx \int_{-\infty}^{\infty} dy \left( \frac{\alpha(T)}{2} P_3^2 + \frac{\beta}{4} P_3^4 + \frac{\delta}{6} P_3^6 + \frac{\zeta}{2} \left( \frac{\partial P_3}{\partial z} \right)^2 + \frac{\eta}{2} \left( \left( \frac{\partial P_3}{\partial y} \right)^2 + \left( \frac{\partial P_3}{\partial x} \right)^2 \right) \right.$$

$$\left. - P_3 \frac{E_3^d}{2} - Q_{ij33} \sigma_{ij} P_3^2 - \frac{s_{ijkl}}{2} \sigma_{ij} \sigma_{kl} \right) + \quad (5)$$

$$+ \int_{-\infty}^{\infty} dx \int_{-\infty}^{\infty} dy \left( \frac{\alpha_0^S}{2} P_3^2(z=0) + \frac{\alpha_h^S}{2} P_3^2(z=h) + \frac{\beta_0^S}{4} P_3^4(z=0) + \frac{\beta_h^S}{4} P_3^4(z=h) + \ldots \right)$$

The gradient terms $\zeta > 0$ and $\eta > 0$, expansion coefficients $\delta > 0$, while $\beta < 0$ for the first order phase transitions or $\beta > 0$ for the second order ones.

Values $\sigma_{jk}$ are the stress tensor components, $Q_{ijkl}$ and $s_{ijkl}$ are the components of electrostriction and elastic compliance tensor correspondingly. If the film thickness is less than the critical thickness for the appearance of misfit dislocations, the free energy coefficients should be renormalized by the elastic stress due to the homogeneous misfit strain between film and substrate.[4] Note that inhomogeneous spontaneous strain always exists in the vicinity of domain wall.[10, 11] Approximate solution for the elastic field of ferroelectric stripe domains in a film on the substrate as a function of misfit strain was given by Stephenson and Elder.[16] Here, we consider the stress due to the inhomogeneity of polarization at the wall-surface junction.

Expansion coefficients $\alpha_{0,h}^S(x,y)$ and $\beta_{0,h}^S(x,y)$ of the surface energy on the polarization powers may be different for different points at the surfaces $z=0$ and $z=h$. The surface energy expansion could be considered for all non-ideal surfaces (e.g. under the presence of different structural defects and intrinsic stresses that determine the order parameter in the "damaged" surface layer). Below we approximate the coordinate dependence by effective values $\alpha_{0,h}^S$ and neglect higher order terms in the surface energy as proposed in Refs.[1, 3, 5, 33, 35 and 36].

Variation of the Gibbs energy (5) on polarization leads to the Euler-Lagrange equation:

$$\left( \alpha - 2 Q_{ij33} \sigma_{ij} \right) P_3 + \beta P_3^3 + \delta P_3^5 - \zeta \frac{\partial^2 P_3}{\partial z^2} - \eta \left( \frac{\partial^2 P_3}{\partial x^2} + \frac{\partial^2 P_3}{\partial y^2} \right) = E_3^d, \quad (6)$$

Boundary conditions for polarization are

$$\left. \left( P_3 - \lambda_1 \frac{\partial P_3}{\partial z} \right) \right|_{z=0} = 0, \quad \left. \left( P_3 + \lambda_2 \frac{\partial P_3}{\partial z} \right) \right|_{z=h} = 0. \quad (7)$$

The introduced extrapolation length $\lambda_{1,2} = \zeta/\alpha_{0,h}^S$ (see Refs.[1, 3, 5, 33, 35]) may be different for $z=0$ and $z=h$. Infinite extrapolation length corresponds to ideal surface ($\alpha_{0,h}^S \to 0$), while zero extrapolation length corresponds $P_3(z=0)=0$ at strongly damaged surface without long-range order. Below we mainly consider the case of equal values $\lambda_{1,2}=\lambda$ for the sake of simplicity. Reported experimental values are $\lambda = 0.5 - 50$nm.[37] It is clear from the boundary conditions (7) that extrapolation length can reflect a different surface polar state that mathematically leads to effects similar to dielectric gap not considered here (see Table 1 in Ref. [38]).

**4. Domain wall surface broadening caused by inhomogeneous elastic stress**

Let us consider the influence of film mechanically free upper surface (upper electrode is absent or regarded very thin) and inhomogeneous elastic stress on a stripe domain structure in uniaxial proper ferroelectrics using perturbation theory. The spatial polarization profile across the domain wall is represented as:

$$P_3(x,z) = P_0(x) + p(x,z). \qquad (8)$$

Function $P_0(x)$ is unperturbed by the film surface influence or "bulk" 1D-domain structure. Perturbation $p(x,z)$ is caused by spatial confinement in z-direction.

The distribution $P_3(x)$ induces inhomogeneous stress. To yield equilibrium stress distribution, free energy (5) is minimized with respect to stress tensor components, $\sigma_{jk}$, as $\partial G/\partial \sigma_{jk} = -\left(Q_{jk33}P_3^2 + s_{jklm}\sigma_{lm}\right) = -u_{jk}$, where $u_{ij}$ is the strain tensor. Additional mechanical equilibrium conditions $\partial \sigma_{ij}(\mathbf{x})/\partial x_i = 0$ and compatibility relation,[39],[11] $e_{ikl}e_{jmn}\left(\partial^2 u_{ln}/\partial x_k \partial x_m\right) = 0$ ($e_{ikl}$ is the permutation symbol or anti-symmetric Levi-Civita tensor[11]), as well as mechanical boundary conditions for zero stress at mechanically free surface $\sigma_{3j}(z=0) = 0$[40] and infinity $\sigma_{ij}(r \to \infty) = 0$ should be satisfied. For the cases of the clamped system with defined displacement components (or with mixed boundary conditions) one should find the equilibrium state as the minimum of the Helmholtz free energy $F = G + \int_V d^3r \cdot u_{jk}\sigma_{jk}$ originating from Legendre transformation of $G$.[4]

Using classical approach,[10, 11] we solved elastic problem with a fixed 1D-distribution of polarization $P_0(x)$ as zero approximation of perturbation theory. Stress convolution with electrostriction $Q_{ij33}\sigma_{ij}(x,z,P_0(x))$, induced by the unperturbed solution $P_0(x)$, for elastically isotropic semi-infinite ferroelectric material has the form:

$$Q_{ij33}\tilde{\sigma}_{ij}(k_1,z) = \begin{pmatrix} \dfrac{(Q_{11}^2+Q_{12}^2)s_{11}-2Q_{12}Q_{11}s_{12}}{(s_{11}^2-s_{12}^2)} - \\ -|k_1|z\exp(-|k_1|z)\dfrac{Q_{11}s_{11}-Q_{12}s_{12}}{s_{11}^2-s_{12}^2}(Q_{11}-Q_{12}) - \\ -\exp(-|k_1|z)\dfrac{Q_{11}s_{11}-Q_{12}s_{12}}{s_{11}^2-s_{12}^2}\left(Q_{11}+Q_{12}\left(1-2\dfrac{s_{12}}{s_{11}}\right)\right) \end{pmatrix} \Delta\tilde{P}_3^2(k_1) \quad (9)$$

Where Voigt notations are used; $\Delta\tilde{P}_3^2(k_1)$ is the Fourier image of the difference $\Delta P_3^2(x)=(P_S^2-P_0^2(x))$, $P_S$ is bulk spontaneous polarization [see subsection A.2 in Appendix A for details]. For the second-order ferroelectrics $P_S^2=-\alpha/\beta$, while $P_S^2=\left(\sqrt{\beta^2-4\alpha\delta}-\beta\right)/2\delta$ for the first order ones. Modified solution (9) $Q_{ij33}(\tilde{\sigma}_{ij}(k_1,z)+\tilde{\sigma}_{ij}(k_1,h-z))$ considered hereinafter is valid for film thickness more than 5-10 longitudinal correlation length of single-domain bulk material $L_z^b$.

Substitution of inhomogeneous stresses (9) into Eq.(6) leads to the following renormalization of coefficients α and β near the surface:

$$\alpha_S(z=0)=\alpha\left(1-2Q_{12}\dfrac{Q_{12}(s_{11}+2s_{12})-Q_{11}s_{11}}{s_{11}(s_{11}+s_{12})\alpha}P_S^2\right), \quad (10a)$$

$$\beta_S(z=0)=\beta\left(1+2Q_{12}\dfrac{Q_{12}(s_{11}+2s_{12})-Q_{11}s_{11}}{s_{11}(s_{11}+s_{12})\beta}\right), \quad (10b)$$

While in the bulk

$$\alpha_S(z\gg L_z^b)=\alpha\left(1-2\dfrac{(Q_{11}^2+Q_{12}^2)s_{11}-2Q_{12}Q_{11}s_{12}}{(s_{11}^2-s_{12}^2)\alpha}P_S^2\right), \quad (11a)$$

$$\beta_S(z\gg L_z^b)=\beta\left(1+2\dfrac{(Q_{11}^2+Q_{12}^2)s_{11}-2Q_{12}Q_{11}s_{12}}{(s_{11}^2-s_{12}^2)\beta}\right). \quad (11b)$$

This immediately leads to different transverse correlation radius near the domain wall. Namely, at the surface

$$L_{\perp,z}(z=0) \approx L_{\perp,z}^b \left(1 + \frac{4Q_{12}(Q_{12}(s_{11}+2s_{12}) - Q_{11}s_{11})}{s_{11}(s_{11}+s_{12})(\alpha + 3\beta P_S^2 + 5\delta P_S^4)} P_S^2\right)^{-1/2}. \quad (12a)$$

While far from the surface:

$$L_{\perp,z}(z \to \infty) \approx L_{\perp,z}^b \left(1 + 4\frac{(Q_{11}^2 + Q_{12}^2)s_{11} - 2Q_{12}Q_{11}s_{12}}{(s_{11}^2 - s_{12}^2)(\alpha + 3\beta P_S^2 + 5\delta P_S^4)} P_S^2\right)^{-1/2}. \quad (12b)$$

Here $L_{\perp}^b = \sqrt{\eta/(\alpha + 3\beta P_S^2 + 5\delta P_S^4)}$ and $L_z^b = \sqrt{\zeta/(\alpha + 3\beta P_S^2 + 5\delta P_S^4)}$ are stress-free transverse and longitudinal correlation length. The stress-free correlation lengths are typically from several to tens of lattice constants below the phase transition but they depend on temperature and tend to infinity at phase transition temperature for the second order ferroelectrics.[41]

Estimations of correlation length in typical ferroelectrics are summarized in Table 1. Striction and free energy expansion coefficients were taken from Refs. [42, 43]. Note that for diffraction methods the observable quantity methods is $L_{\perp,z}$, not $L_{\perp,z}^b$. Similar mechanism of elastic stress influence on domain wall width should exist in all ferroic materials.

**Table 1.** Dielectric permittivity $\varepsilon_{ii}$ and correlation radii ratio for typical ferroelectrics.

| Material | $\varepsilon_{11}$ | $\varepsilon_{33}$ | $L_{\perp,z}(0)/L_{\perp,z}^b$ | $L_{\perp,z}(\infty)/L_{\perp,z}^b$ | $L_{\perp,z}(0)/L_{\perp,z}(\infty)$ |
|---|---|---|---|---|---|
| PbZr$_{0.6}$Ti$_{0.4}$O$_3$ | 529 | 295 | 0.63 | 0.54 | 1.17 |
| PbZr$_{0.5}$Ti$_{0.5}$O$_3$ | 1721 | 382 | 0.30 | 0.27 | 1.11 |
| PbZr$_{0.4}$Ti$_{0.6}$O$_3$ | 498 | 197 | 0.36 | 0.28 | 1.27 |
| PbTiO$_3$ | 140 | 105 | 0.66 | 0.58 | 1.14 |
| BaTiO$_3$ | 2920 | 168 | 0.80 | 0.74 | 1.09 |
| LiNbO$_3$ | 85 | 29 | 0.996 | 0.986 | 1.01 |
| LiTaO$_3$ | 54 | 44 | 0.994 | 0.988 | 1.01 |

The ratios $L_{\perp,z}(0)/L_{\perp,z}(\infty)$ can be closer to 1 allowing for the stress relaxation on the defects typically concentrated in the vicinity of domain walls. Qualitatively, inhomogeneous elastic stress leads to essential clamping and contraction of domain wall width in perovskites with high striction coefficients, since the wall width $w(z) \sim L_{\perp}(z)$, and it slightly increases when approaching the surface because $L_\perp(0)/L_\perp(\infty) > 1$. However we obtained the ratio $1 < L_\perp(0)/L_\perp(\infty) < 1.3$ and so $1 < w(0)/w(\infty) < 1.3$ allowing for inhomogeneous elastic stress effect only for materials from Tab.1, while recent experimental results obtained by means of scanning nonlinear dielectric microscopy[44, 45] reveal higher ratio $2 < w(0)/w(\infty) < 5$ for LiTaO$_3$. Extra broad domain walls with $w(0) \sim 100$ nm were observed at LiNbO$_3$ surface,[46] where we estimated $L_{\perp,z}(0)/L_{\perp,z}(\infty) \approx 1$ and $L_{\perp,z} \approx L_{\perp,z}^b$. These facts force us to consider other mechanisms for surface-induced domain wall broadening, such as finite extrapolation length effect.

It should be noted, that similar perturbation approach was applied to the consideration of domain structure in ferroelastic films on the substrate.[26] There exists an alternative approach, utilizing the fictitious compensating forces on the free surfaces and interphase boundaries, applied to provide lattice matching between different phases and substrate at fixed values of order parameter (self-strain).[47] As it was shown in Ref. [27], the first order analytical solution of perturbation method was not able to reproduce the general features of numerical solution of coupled problem, while the method of compensating surface forces gave better results. However, one can hardly project the results of Lee et al.[27] on the considered problem, since we do not consider the coupling (electrostriction) coefficients as a small parameter and do not assume that the solution is separable, as Lee et al did. Furthermore, the effect of striction on wall thickness in ferroelectric perovskites is not weak, but the difference between the surface and bulk is not so pronounced (see Table I). The latter justifies using bulk renormalization of free energy coefficients in the subsequent consideration.

## 5. Domain wall broadening caused by finite extrapolation length

Now let us consider the single domain wall width due to a finite extrapolation length in the $z$-direction. For the first order ferroelectrics the single domain wall bulk profile is

$$P_0(x) = \frac{P_S \cdot \sinh\left((x-x_0)/2L_\perp\right)}{\sqrt{3(2\alpha_S + \beta_S P_S^2)/(4\alpha_S + \beta_S P_S^2) + \sinh^2\left((x-x_0)/2L_\perp\right)}}$$

(see e.g. Refs. [11, 16]), while it reduces to $P_0(x) = P_S \tanh\left((x-x_0)/2L_\perp\right)$ for the second order ones.[10] Correlation radius $L_\perp$ is given by Eq.(12b), the wall plane is $x = x_0$.

In Appendix B we obtained the trial function for the free energy functional (5):

$$\tilde{P}_3(\mathbf{k},z) = \tilde{P}_0(\mathbf{k})\left(1 + P_V\begin{pmatrix} A(s_1,s_2)(\cosh(s_1 z) + \cosh(s_1(h-z))) \\ + A(s_2,s_1)(\cosh(s_2 z) + \cosh(s_2(h-z))) \end{pmatrix}\right) \quad (13)$$

Amplitude $P_V$ is a variational parameter, coordinate dependent part is the solution of Eq.(3a) linearized allowing substitution (8). For equal extrapolation length $\lambda_{1,2}=\lambda$ we derived that

$$A(s,q) = \frac{\sinh(qh/2)M(q)}{\cosh(sh/2)Det_I(s,q,\lambda,h)}, \quad (14)$$

Where function $M(s) = \dfrac{s}{\varepsilon_{11}k^2 - s^2}$. System determinant is given by:

$$Det_I(s,q,\lambda,h) = 2\begin{pmatrix} M(s)\cosh\left(\dfrac{qh}{2}\right)\sinh\left(\dfrac{sh}{2}\right) - M(q)\cosh\left(\dfrac{sh}{2}\right)\sinh\left(\dfrac{qh}{2}\right) + \\ + \lambda(s M(q) - q M(s))\sinh\left(\dfrac{sh}{2}\right)\sinh\left(\dfrac{qh}{2}\right) \end{pmatrix} \quad (15)$$

Values $s_1(k)$ and $s_2(k)$ are the roots of the characteristic equation:

$$(s^2 - k^2\varepsilon_{11})(1 - (L_z^2 s^2 - L_\perp^2 k^2)) = -(\varepsilon_{33} - 1)s^2. \quad (16)$$

Hereinafter $k = \sqrt{k_1^2 + k_2^2}$, at that permittivity $\varepsilon_{33} = 1 + \dfrac{1}{\varepsilon_0(\alpha_S + 3\beta_S P_S^2 + 5\delta P_S^4)}$ and correlation lengths $L_{\perp,z}$ are given by Eqs.(12) neglecting small difference between $L_{\perp,z}(0)$ and $L_{\perp,z}(\infty)$.[48]

In Appendix B we show, that determinant $Det_I$ given by Eq.(15) can be zero at several negative $\lambda$ values. The case of zero $Det_I$ could not be treated in terms of perturbation approach (13) based on the bulk domain structure $P_0(x)$. Thus below we consider the range of positive extrapolation length values ($\lambda>0$) and high enough negative $\lambda \leq -2$, where the bulk domain structure $P_0(x)$ is stable and thus $P_V \approx 1$ is the good first approximation for film thickness more than 5-10 critical thickness defined below.

For particular case $\mathbf{k} \to 0$, characteristic values are $s_1^2 \approx k^2 \varepsilon_{11}/\varepsilon_{33} \to 0$, $s_2^2 \approx \varepsilon_{33}/L_z^2 \approx l_z^{-2}$, where the thickness $l_z = \sqrt{\zeta \varepsilon_0} \leq 2\text{A}^0$. Under the condition $h/l_z \gg 1$, we obtained from Eq.(13) the following renormalization of spontaneous polarization $\tilde{P}_3(0,z) \approx P_S\left(1 + \dfrac{2\varepsilon_{33}l_z}{(1+\lambda/l_z)h}\right)^{-1}\left(1 - \dfrac{\exp(z/l_z) + \exp((h-z)/l_z)}{1 + (\lambda/l_z)}\right)$. The expression coincides with the polarization distribution considered in Ref.[34] for transversally homogeneous film as anticipated. Corresponding solution is approximately constant outside the ultrathin layer of thickness $z$ about several $l_z$, while spontaneous polarization decrease in comparison with the bulk value $P_S$ becomes negligible for thickness $h \gg h_{cr}$, where the critical thickness $h_{cr} \approx 4\varepsilon_{33} l_z / (1 + \lambda/l_z)$ is not more than 5-10 nm at $\lambda = 0.5...50$ nm and typical material parameters. For the thickness $h \gg h_{cr}$ all our results obtained for bulk values $P_S$, $\varepsilon_{33}$ and $P_V=1$ are self-consistent. The Fourier image of polarization distribution (13) for the case of film thickness $h \gg h_{cr}$ can be simplified as:

$$\tilde{P}_3(\mathbf{k},z) \approx \tilde{P}_0(\mathbf{k}) \left( \begin{array}{l} 1 - \dfrac{(\exp(-s_1 z) + \exp(-s_1(h-z)))(\varepsilon_{11} k^2 - s_1^2) s_2}{(s_2 - s_1)(\varepsilon_{11} k^2 + s_1 s_2 + \lambda s_1 s_2 (s_1 + s_2))} + \\ + \dfrac{(\exp(-s_2 z) + \exp(-s_2(h-z)))(\varepsilon_{11} k^2 - s_2^2) s_1}{(s_2 - s_1)(\varepsilon_{11} k^2 + s_1 s_2 + \lambda s_1 s_2 (s_1 + s_2))} \end{array} \right). \quad (17)$$

In the framework of linear imaging theory, the spectral ratio $W(\mathbf{k},z) = \tilde{P}_3(\mathbf{k},z) / \tilde{P}_0(\mathbf{k})$ is the transfer function of domain structure $P_0(x)$ generated by the lateral confinement conditions. It is clear from Figs. 2-3 that $W(\mathbf{k},z)$ decreases when approaching the surface for positive and increases for negative extrapolation length, $\lambda$ (compare Figs.2 (a) with (b)). Correspondingly, the domain wall broadens at the surface for positive and contracts for negative $\lambda$. We used expressions (13)-(16) and inverse Fourier transformation when calculating the curves in Figs.2-3.

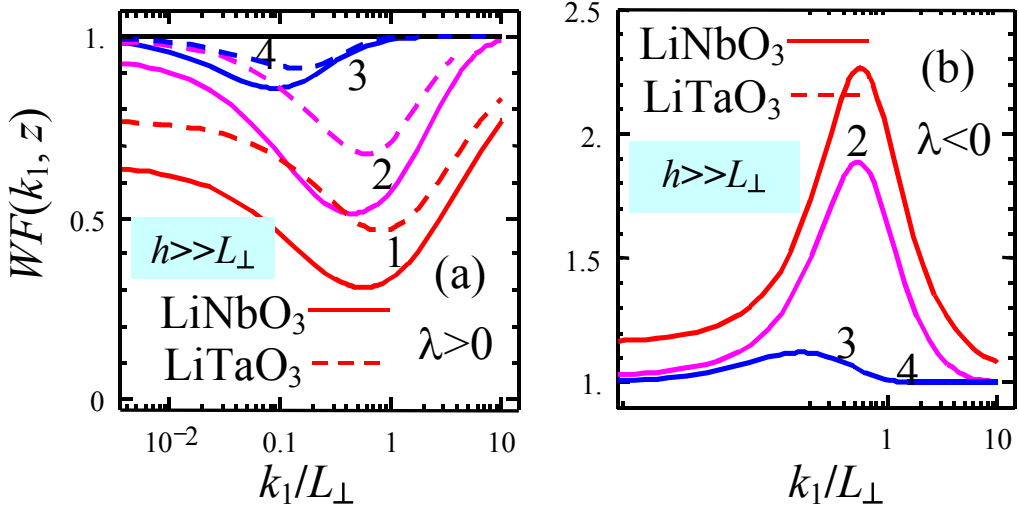

**Figure 2.** (Color online) Domain wall spectra transfer function $W(k,z)$ calculated from Eq.(17) for a thick ($h \gg L_\perp$) LiNbO$_3$ sample ($\varepsilon_{11}=84$, $\varepsilon_{33}=30$, $L_z/L_\perp=1.5$, solid curves) and LiTaO$_3$

($\varepsilon_{11}=54$, $\varepsilon_{33}=44$, $L_z/L_\perp=1$, dashed curves) for positive extrapolation length $\lambda/L_\perp=0.5$ (a) and negative $\lambda/L_\perp=-2$ (b) at different distances $z$ from surface $z/L_\perp=0, 0.5, 5, \infty$ (curves 1, 2, 3, 4).

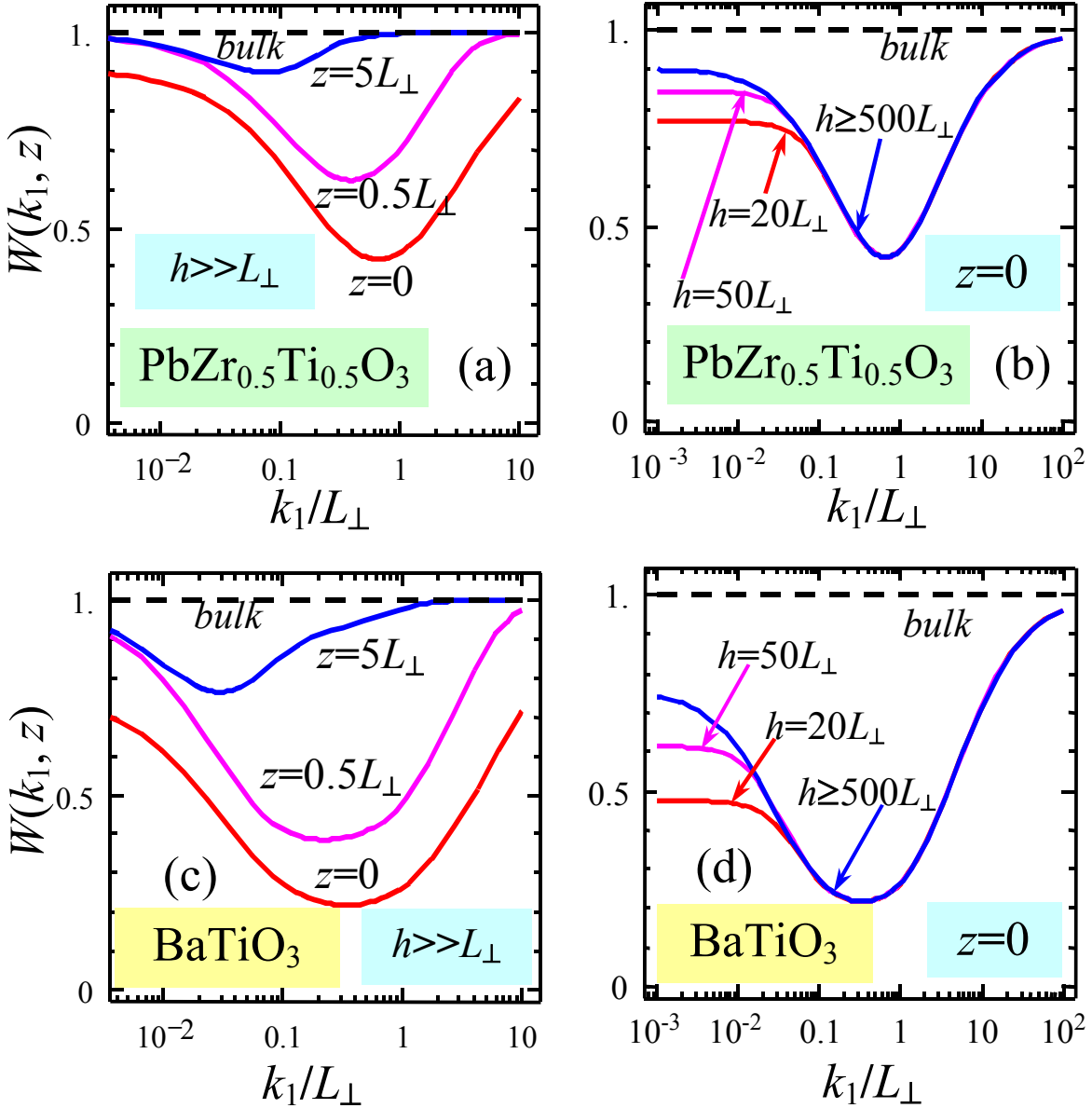

**Figure 3.** (Color online) Domain wall spectra transfer function $W(k,z)$ calculated from Eq.(13) at extrapolation length $\lambda=0.5L_\perp$ for material parameters of PbZr$_{0.5}$Ti$_{0.5}$O$_3$ ($L_z=L_\perp$, $\varepsilon_{11}=\varepsilon_{33}=507$) and BaTiO$_3$ ($L_z=2L_\perp$, $\varepsilon_{11}=1200$, $\varepsilon_{33}=80$). Curves in plots (a,c) correspond to different distances $z$ from the surface $z/L_\perp=0, 0.5, 5, \infty$ of thick film ($h \gg L_z$); curves in plots (b,d) correspond to the surface $z=0$ of film with different thickness $h$ (see labels near the curves).

At the surface of films with different thickness, corresponding transfer functions differ only at small $k_1$ values, at that the small-$k$ plateau value decreases with film thickness decrease [compare curves in Figs.3(b,d) for various $h$].

Analyzing results presented in Figs.2-3 for $k$-space proving that $W(k,z)$ decreases when approaching the film surface for positive extrapolation length $\lambda$, one could expect opposite situation for the single domain wall $x$-profile: domain wall should broaden when approaching the surface for positive extrapolation length $\lambda$ and contracts for negative ones. Moreover, coincidence of transfer functions at high-$k$ should lead to the same $x$-profile in the immediate vicinity of domain wall (i.e. at $|x| < L_\perp$), while different plateau at small-$k$ should lead to the different saturation rate of $x$-profile far from the wall.

Actually, it is clear from Figs. 4 that a single domain wall broadens when approaching the surface $z=0$ for positive $\lambda$ and contracts for negative ones. Figs.5 demonstrate that domain wall slope is almost independent on film thickness $h$ at film surface $z=0$, while it is essentially differ from the bulk profile [compare curves in linear and log-linear scale].

Calculated width of domain wall $w$ at level 0.76 as a function of depth $z$ from the sample surface is shown in Figs.6a,b for $PbZr_{0.5}Ti_{0.5}O_3$ material parameters and equal extrapolation lengths $\lambda$ for both film surfaces. We used expressions (13)-(16) and inverse Fourier transformation when generating these plots. It is evident that smaller extrapolation length leads to the strongest broadening [compare plots (a) and (b)].

Calculated width (solid curves) of domain wall at level 0.76 as a function of its depth from the surface of $LiTaO_3$ is shown in Figs.6c in comparison with experimental data[45] in 500nm thick stoichiometric $LiTaO_3$ (squires) and 50nm thick congruent $LiTaO_3$ (triangles). When calculating the curve for 50nm thick $LiTaO_3$ film we taking into account that domain wall profile $w$ is strongly asymmetric, namely at the surface $z=0$ the width is 5 times bigger than saturated "bulk" value near the surface $z=h$. Therefore we conclude that extrapolation length $\lambda_2(h) \gg \lambda_1(0)$. For the case $\lambda_2 \to \infty$ domain wall broadening is essential only near the surface $z=0$ (where $\lambda_1$ is finite), while the surface $z=h$ is indistinguishable from the bulk. Thus for $\lambda_2 \to \infty$, one should use expressions (17) for profile calculations inside the film ($0 \le z \le h$) after the substitution of double thickness $2h$.[49] Dotted curves in Fig.6c are numerical calculations by using phase field method [50] (see Appendix C for details). It is clear from Fig.6c, that analytical calculations are in a reasonable agreement with experimental data and numerical simulations. The presence of damaged surface layer, reported in Ref.[45], and not-measured surface polarization value allow us to consider finite extrapolation length value as a fitting parameter.

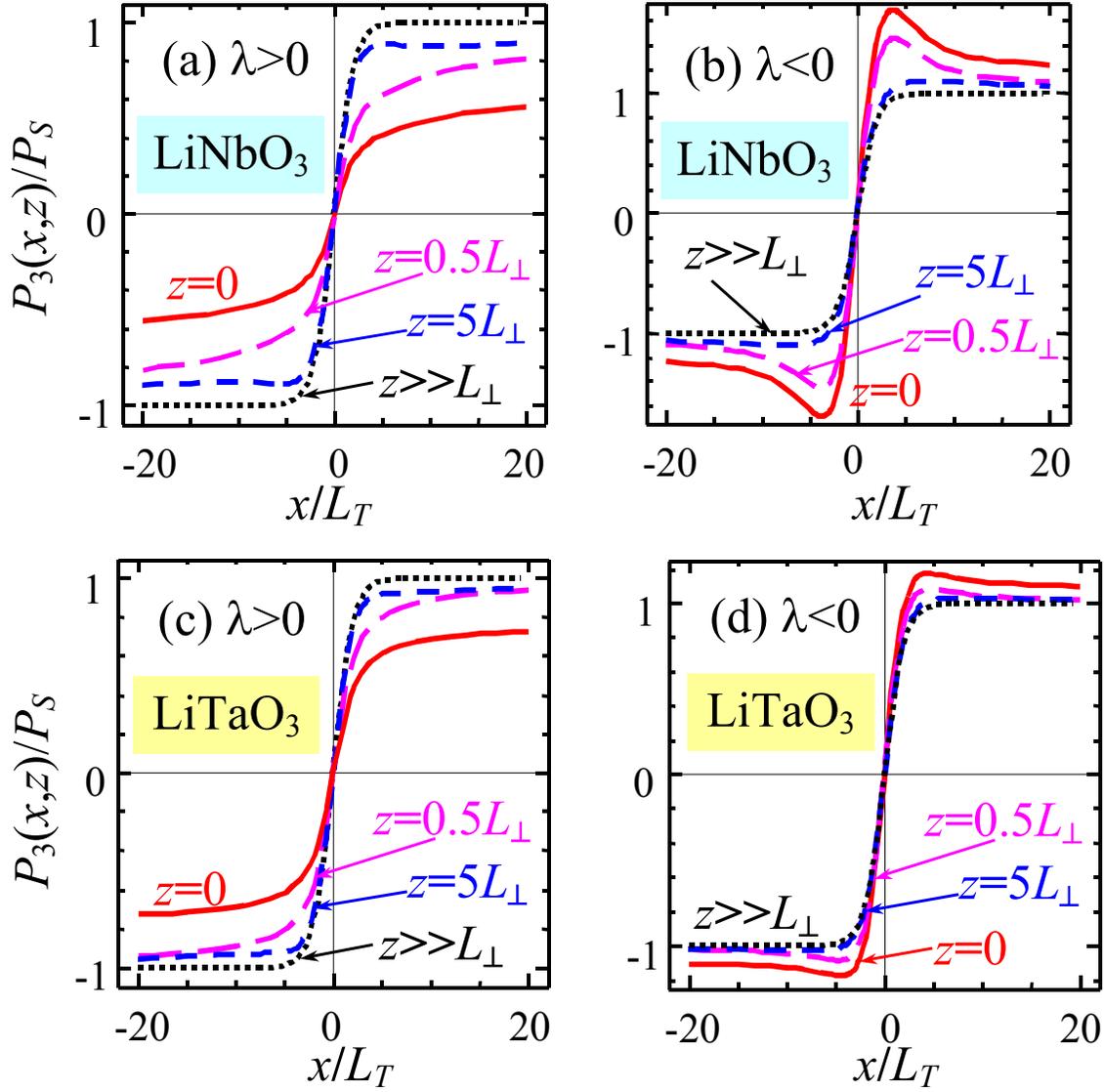

**Figure 4.** (Color online) Normalized domain wall profile $P_3(x,z)/P_S$ in (a,b) LiNbO$_3$ ($\varepsilon_{11}$=84, $\varepsilon_{33}$=30, $L_z/L_\perp$=1.5) and (c,d) LiTaO$_3$ ($\varepsilon_{11}$=54, $\varepsilon_{33}$=44, $L_z/L_\perp$=1), for positive extrapolation length $\lambda/L_\perp$=0.5 (a,c) and negative $\lambda/L_\perp$=-2 (b,d) at different distances $z$ from surface $z/L_\perp$=0, 0.5, 5, ∞ (see labels near the curves).

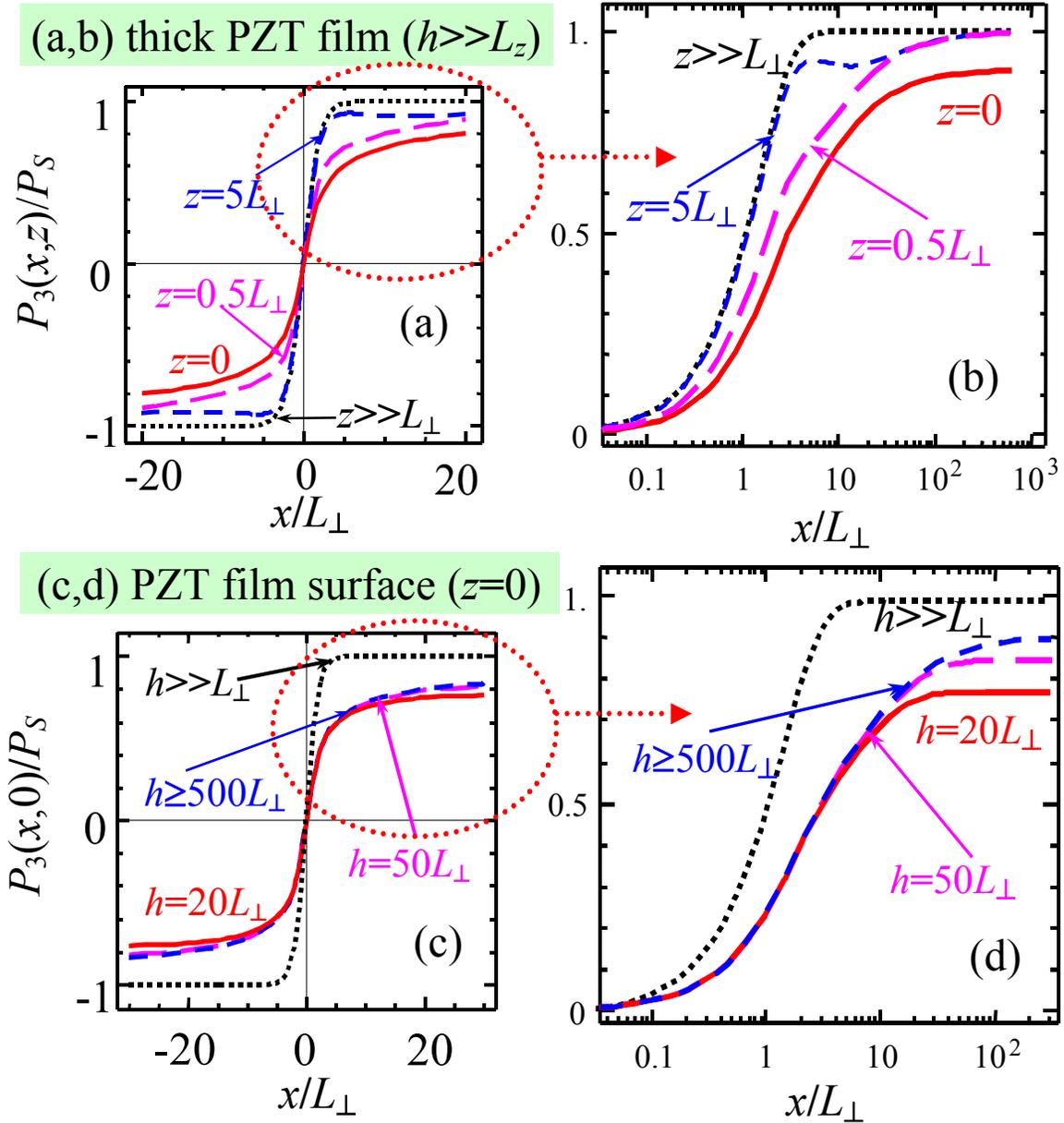

**Figure 5.** (Color online) Normalized domain wall profile $P_3(x,z)/P_S$ calculated in PbZr$_{0.5}$Ti$_{0.5}$O$_3$ films for $L_z=L_\perp$ and extrapolation length $\lambda=0.5L_\perp$. Profiles in linear scale are shown in (a,c), half-profiles in log-linear scale are shown in (b,d). Curves in plots (a,b) correspond to different distances $z/L_\perp=0$, 0.5, 5, $\infty$ from the surface of thick film ($h\gg L_z$). Curves in plot (c,d) correspond to the surface $z=0$ of film with different thickness $h$ (see labels near the curves).

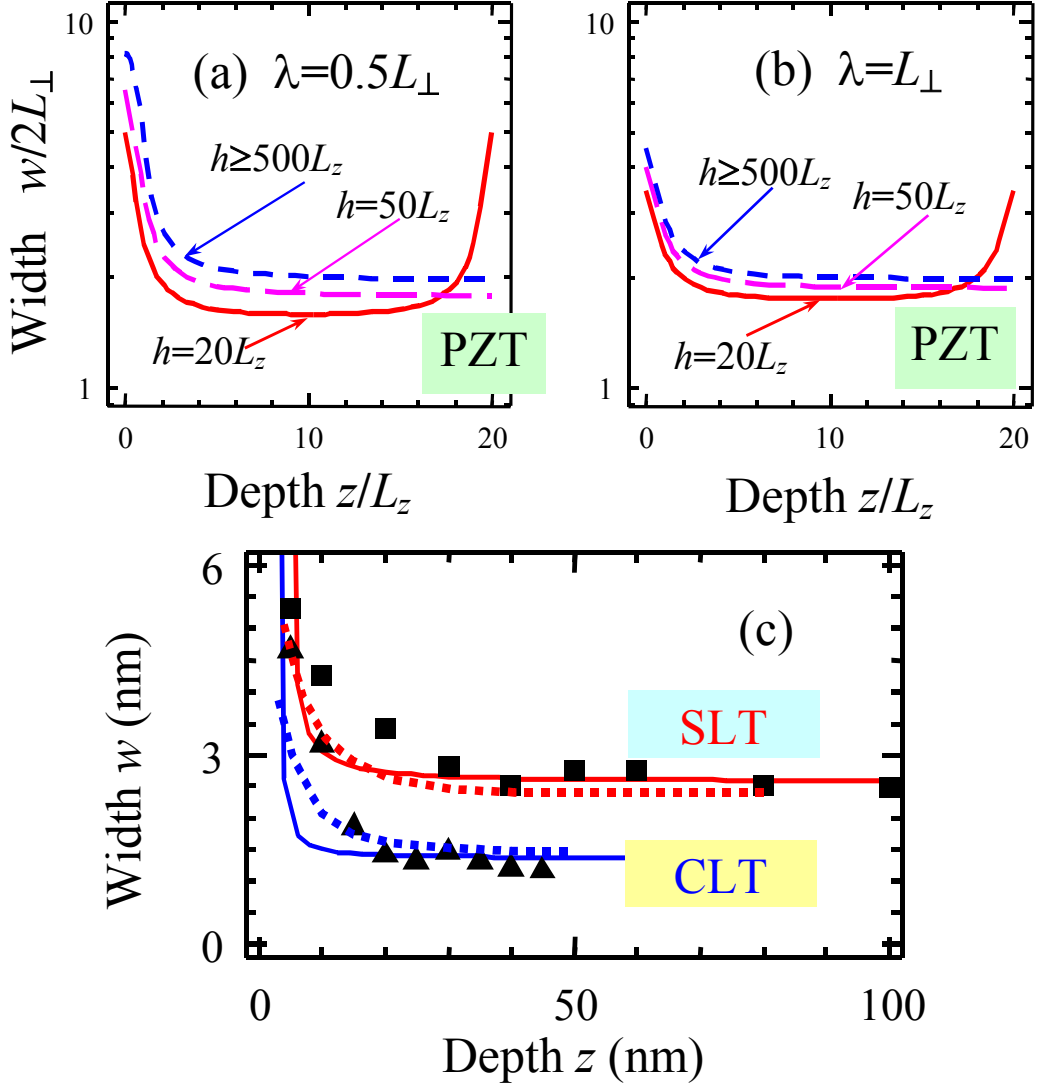

**Figure 6.** (Color online) (a,b) Thickness of domain wall $w/2L_\perp$ at level 0.76 as a function of depth $z$ from the surface at $L_z=L_\perp$, extrapolation length $\lambda=0.5L_\perp$ (a) and $\lambda=L_\perp$ (b) for material parameters of $PbZr_{0.5}Ti_{0.5}O_3$ (PZT). (c) Thickness of domain wall at level 0.76 as a function of its depth from the surface of $LiTaO_3$. Squires are experimental data from Refs.[44],[45] for 500nm thick stoichiometric $LiTaO_3$ (SLT), triangles correspond to 50nm thick congruent $LiTaO_3$ (CLT). Solid curves are analytical calculations based on Eqs.(13)-(16) for fitting parameters $L_\perp=1.3$ nm, $L_z=1.6$ nm, different extrapolation lengths $\lambda_1(0)=0.1$ nm and ($\lambda_2(h)$ value appeared not important), for SLT; while $L_\perp=0.7$ nm, $L_z=1.4$ nm and $\lambda_1(0)=0.1$ nm, $\lambda_2(h)\gg 30$nm for CLT. Corresponding dotted curves are numerical calculations by phase field modeling for the same fitting parameters.

To corroborate analytically the domain wall surface broadening caused by finite extrapolation length, let us analyze the Fourier image of polarization distribution for semi-infinite sample ($h \to \infty$). Expanding Eq.(17) over $k$ value as

$$\tilde{P}_3 \approx \tilde{P}_0(\mathbf{k}) - \frac{\tilde{P}_0(\mathbf{k})\exp(-\sqrt{\varepsilon_{33}}\,z/L_z) + \exp(-kz\sqrt{\varepsilon_{11}/\varepsilon_{33}})\sqrt{\varepsilon_{11}}L_z \cdot k}{1 + \sqrt{\varepsilon_{33}}(\lambda/L_z) + \left(\sqrt{\varepsilon_{11}}L_z + \lambda\sqrt{\varepsilon_{11}/\varepsilon_{33}}\right)k + \left(L_\perp^2 + \left(\sqrt{\varepsilon_{33}}L_\perp^2/L_z + L_z\varepsilon_{11}/\sqrt{\varepsilon_{33}}\right)\lambda\right)k^2/2}, \quad (18)$$

we obtain the approximate expression for original determination:

$$P_3(x,z) \approx \left(1 - \frac{\exp(-\sqrt{\varepsilon_{33}}\,z/L_z)}{1 + \lambda\sqrt{\varepsilon_{33}}/L_z}\right)P_0(x) - \frac{2\sqrt{\varepsilon_{11}}L_z}{1 + \lambda\sqrt{\varepsilon_{33}}/L_z}\int_{-\infty}^{\infty} dy\, P_0(y)\Delta(x - x_0 - y, z), \quad (19)$$

Where the function $\Delta(x,z) = \dfrac{(\varepsilon_{11}/\varepsilon_{33})(z + L_z^2/(L_z + \lambda))^2 - x^2}{\left(x^2 + (\varepsilon_{11}/\varepsilon_{33})(z + L_z^2/(L_z + \lambda))^2\right)^2}$. Since $\varepsilon_{33} \gg 1$ as well as $L_z/\sqrt{\varepsilon_{33}} \leq 2\text{Å}^0$, the terms proportional to $\exp(-\sqrt{\varepsilon_{33}}\,z/L_z)$ vanish with distance $z$ increase much quicker then the other ones. So, the convolution behavior in the second term determines the change of domain wall width $w(z)$ caused by the surface. For the second order ferroelectrics direct integration in Eq.(19) leads to the approximate analytical expression

$$P_3(x,z) \approx \left(1 - \frac{\exp(-\sqrt{\varepsilon_{33}}\,z/L_z)}{1 + \lambda\sqrt{\varepsilon_{33}}/L_z}\right)P_S \tanh\left(\frac{x - x_0}{2L_\perp}\right) -$$
$$- \frac{\sqrt{\varepsilon_{11}}\,P_S\,L_z/L_\perp}{4(1 + \lambda\sqrt{\varepsilon_{33}}/L_z)}\ln\left(\frac{(2L_\perp + x - x_0)^2 + (\varepsilon_{11}/\varepsilon_{33})(z + L_z^2/(L_z + \lambda))^2}{(2L_\perp - x + x_0)^2 + (\varepsilon_{11}/\varepsilon_{33})(z + L_z^2/(L_z + \lambda))^2}\right) \quad (20)$$

It is clear from Eq.(20), that finite extrapolation length effect provides power saturation of domain wall profile (the last logarithmic term should be expanded far from the wall), that is much slower compared to exponential saturation of bulk profile $P_0(x)$. Under the condition $L_\perp \approx L_z$, the amplitude of the second term is proportional to $\sqrt{\varepsilon_{11}/\varepsilon_{33}}\,L_{\perp,z}/\lambda$. So, the smaller the ratio $L_{\perp,z}/\lambda$ the strongest is the surface domain wall broadening.

At distances $|x - x_0| \gg L_\perp$ the latter term behaves as $2(x - x_0)/\left((x - x_0)^2 + (\varepsilon_{11}/\varepsilon_{33})(z + L_z^2/(L_z + \lambda))^2\right)$, which is the distribution of stray depolarization field far from the break of double electric layer.[51] The term $R_z^2/(L_z + \lambda)$ plays the role of effective double layer (screening layer) width (see also Table 1 in Ref. [38]). Thus the expression (9) explains power saturation of profiles (7)-(8) as the direct effect of the depolarization field which decreases much slowly in comparison with exponential saturation of bulk profile "$\tanh((x-x_0)/2L_\perp)$".

Qualitatively the same effect of finite extrapolation length and inhomogeneous elastic stress on periodic 180°-domain structure near the film surface was obtained numerically by using phase field method [see dotted curves in Figs.6c, Figs.7 and Appendix C].

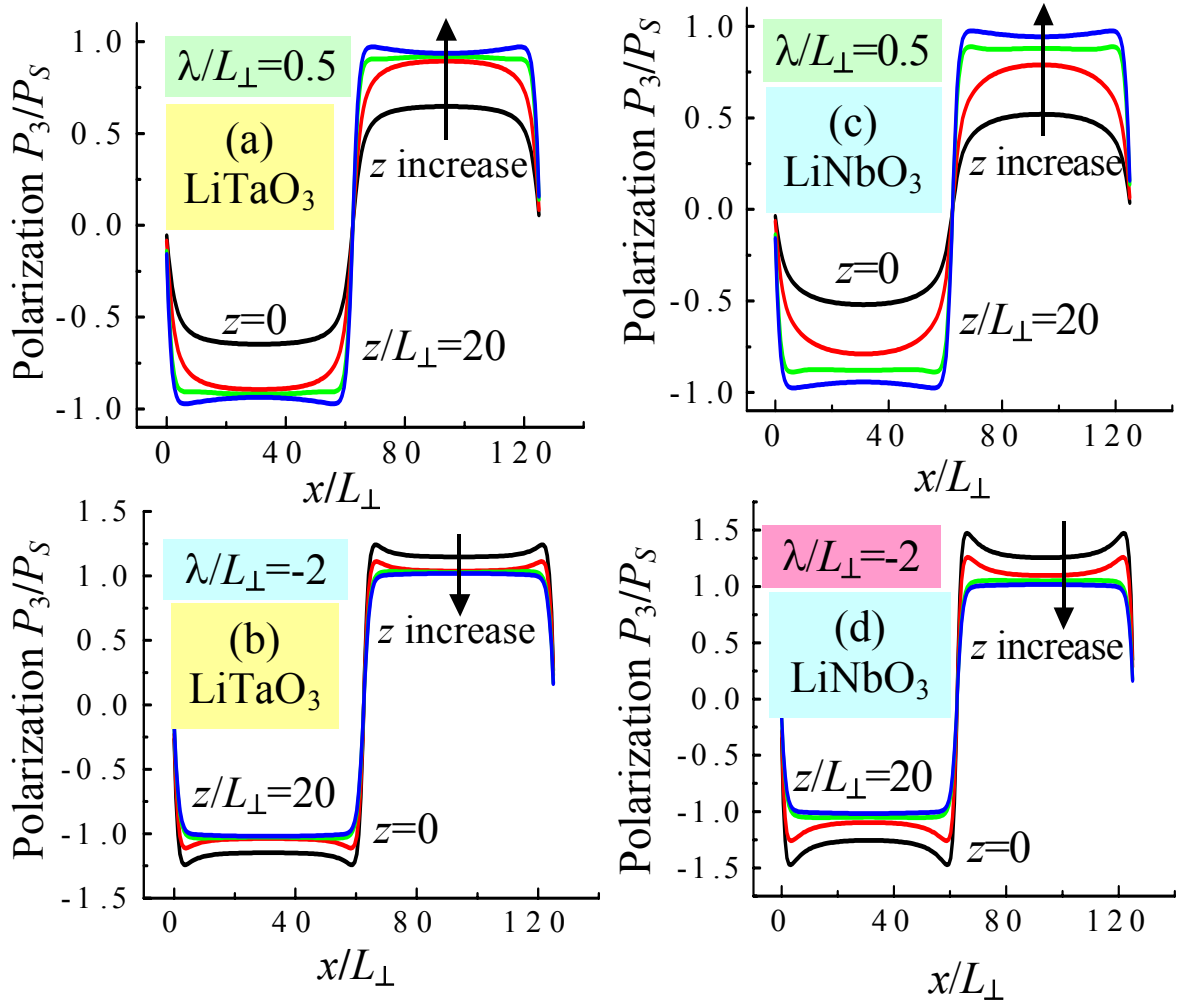

**Figure 7.** (Color online). Domain structure near the surface of LiTaO$_3$ (a), (b) and LiNbO3 (c), (d) films calculated numerically by using phase field method for $L_z=L_\perp$ (a,b) and $L_z=1.5L_\perp$ (c,d), extrapolation length $\lambda=0.5L_\perp$ (a, c) $\lambda=-2L_\perp$ (b, d) and different distance from the surface $z=0$, $0.5L_\perp$, $5L_\perp$, $20L_\perp$.

To summarize the Sections 4-5, three factors considered here: finite extrapolation length (reflecting the surface energy contribution), spatially dependent depolarization field and inhomogeneous elastic stress determine the inhomogeneity of ferroelectric polarization (and thus the domain wall width) near the sample surface. For ferroelectrics with weak ($LiNbO_3$, $LiTaO_3$) or moderate ($BaTiO_3$ or $PbZr_{0.5}Ti_{0.5}O_3$) piezoelectric coupling (electrostriction constants) extrapolation lengths smaller that correlation length affects the wall width much stronger than inhomogeneous elastic stress, while in materials with high piezoelectric coupling and small surface energy extrapolation length both effects can be comparable. For the case the domain wall profile saturation becomes power when approaching the surface, while it exponentially saturating in the bulk. For extrapolation length much higher than correlation one inhomogeneous elastic stress strongly dominates.

**Conclusion**

We analyze the polarization behavior and domain wall broadening on the ferroelectric wall-surface junction. We demonstrate that even when an electrode or ambient screening minimizes depolarization field, inhomogeneous elastic stress and positive extrapolation length lead to the domain wall broadening. For ferroelectrics with weak piezoelectric coupling (e.g. $LiNbO_3$ or $LiTaO_3$ with small electrostriction constants) extrapolation length affects the wall width much more strongly than inhomogeneous elastic stress, while in materials with high piezoelectric coupling both effects may be comparable. Notably, the wall profile follows a long-range power-law profile at the surface, as opposed to an exponential saturation of the order parameter in the bulk. The saturation law is explained by the behavior of stray depolarization field that decreases like the field created by the break of double electric layer. Note, that effects of domain wall structure changes caused by demagnetization field can be observed in ferromagnetic films, when the Bloch type domain wall transforms into a Neel wall with thickness decrease.[52]

These results have broad implication for fundamental issues such as maximal information storage density in ferroelectric data storage, domain wall pinning mechanisms at surfaces and interfaces, and nucleation dynamics. Furthermore, the quasi-phase matching of optical/acoustical waves in periodically poled ferroelectric media of high order harmonics generators/converters seems to be very sensitive to the fine details of polarization distribution (reproducibility of periodic structure should be about 10 nm). Any long-range depth distribution may result in undesirable dispersion of wave propagation.[53]

**Acknowledgement**

Research was partially (EAE and MDG) supported by Science & Technology Center in Ukraine, project No. 3306. SVK is sponsored by the Oak Ridge National Laboratory (ORNL), managed by UT-Battelle, LLC for the U. S. Department of Energy under Contract No. DE-AC05-00OR22725. VG wishes to gratefully acknowledge financial support from the National Science Foundation grant numbers DMR-0602986, 0512165, 0507146, and 0213623, and CNMS at Oak Ridge National Laboratory. LQ and YL are supported by DOE under grant number DE-FG02-07ER46417 and Los Alamos National Laboratory.

*Supplementary materials to*

**Surface Effect on Domain Wall Width in Ferroelectrics**

Eugene A. Eliseev,[1, §, ††] Anna N. Morozovska,[1, **, ††] Sergei V. Kalinin,[2] Yulan L. Li,[3] Jie Shen,[4] Maya D. Glinchuk,[1] Long-Qing Chen,[3] and Venkatraman Gopalan[3]

[1] Institute for Problems of Materials Science, National Academy of Science of Ukraine, 3, Krjijanovskogo, 03142 Kiev, Ukraine

[2] The Center for Nanophase Materials Sciences and Materials Science and Technology Division, Oak Ridge National Laboratory, Oak Ridge Tennessee 37831, USA

[3] Department of Materials Science and Engineering, Pennsylvania State University, University Park, Pennsylvania 16802, USA

[4] Department of Mathematics, Purdue University, West Lafayette, Indiana 47907, USA

**Appendix A.**

**A.1. Electrostatic potential and depolarization field**

Fourier transformation $f(x,y,z) = \frac{1}{2\pi}\int_{-\infty}^{\infty} dk_1 \int_{-\infty}^{\infty} dk_2 \exp(-ik_1 x - ik_2 y) \cdot \tilde{f}(k_1, k_2, z)$ over $x$ and $y$ coordinates in Eqs.(5), one obtain the differential equation for the image:

$$\frac{d^2 \tilde{\varphi}}{dz^2} - \frac{k^2}{\tilde{\gamma}^2}\tilde{\varphi} = \frac{\tilde{\phi}(\mathbf{k},z)}{\varepsilon_0} \qquad (A.1a)$$

$$\tilde{\varphi}(\mathbf{k},0) = \tilde{\varphi}_e(\mathbf{k}), \quad \tilde{\varphi}(\mathbf{k},h) = 0, \qquad (A.1b)$$

---

[§] E-mail: eliseev@i.com.ua

[**] E-mail: morozo@i.com.ua, permanent address: V.Lashkaryov Institute of Semiconductor Physics, National Academy of Science of Ukraine, 41, pr. Nauki, 03028 Kiev, Ukraine

[††] These authors contributed equally to this work.

Here $\tilde{\gamma} = \sqrt{1/\varepsilon_{11}}$ and $\tilde{\phi}(\mathbf{k},z) = \dfrac{\partial}{\partial z}\tilde{P}_3(\mathbf{k},z)$. Let us find the homogeneous solution of (A.1a) for $\tilde{\varphi}_i(k,z)$ in the form $\tilde{\varphi}_{hom}(k,z) = A(k_1,k_2)\exp(+k\,z/\tilde{\gamma}) + B(k_1,k_2)\exp(-k\,z/\tilde{\gamma})$. Substituting into the boundary conditions (A.1b) leads to:

$$\tilde{\varphi}_{hom}(k,z) = \tilde{\varphi}_e(\mathbf{k})\frac{\exp(-k\,z/\tilde{\gamma}) - \exp(-k(2h-z)/\tilde{\gamma})}{1-\exp(-2k\,h/\tilde{\gamma})}, \tag{A.2}$$

Here $k \equiv \sqrt{k_1^2 + k_2^2}$. The inhomogeneous solution of (A.1a) with homogeneous boundary conditions has the form [54]:

$$\tilde{\varphi}_{part}(k,z) = -\left(\begin{array}{l}\displaystyle\int_0^z d\xi\,\frac{\tilde{\phi}(\mathbf{k},\xi)}{\varepsilon_0\tilde{\gamma}}\frac{\sinh(k\xi/\tilde{\gamma})\sinh(k(h-z)/\tilde{\gamma})}{k\sinh(k\,h/\tilde{\gamma})} + \\ \displaystyle\int_z^h d\xi\,\frac{\tilde{\phi}(\mathbf{k},\xi)}{\varepsilon_0\tilde{\gamma}}\frac{\sinh(k\,z/\tilde{\gamma})\sinh(k(h-\xi)/\tilde{\gamma})}{k\sinh(k\,h/\tilde{\gamma})}\end{array}\right) \tag{A.3}$$

General solution has the form:

$$\tilde{\varphi}(\mathbf{k},z) = \tilde{\varphi}_e(\mathbf{k})\frac{\sinh(k(h-z)/\tilde{\gamma})}{\sinh(k\,h/\tilde{\gamma})} - \left(\begin{array}{l}\displaystyle\int_0^z d\xi\,\frac{\tilde{\phi}(\mathbf{k},\xi)}{\varepsilon_0\tilde{\gamma}}\frac{\sinh(k\xi/\tilde{\gamma})\sinh(k(h-z)/\tilde{\gamma})}{k\sinh(k\,h/\tilde{\gamma})} + \\ \displaystyle\int_z^h d\xi\,\frac{\tilde{\phi}(\mathbf{k},\xi)}{\varepsilon_0\tilde{\gamma}}\frac{\sinh(k\,z/\tilde{\gamma})\sinh(k(h-\xi)/\tilde{\gamma})}{k\sinh(k\,h/\tilde{\gamma})}\end{array}\right) \tag{A.4}$$

Here $\tilde{\phi}(\mathbf{k},z) = \dfrac{\partial}{\partial z}\tilde{P}_3(\mathbf{k},z)$, and so integration over parts in Eq.(A.4) leads to the Fourier representation of electrostatic potential:

$$\tilde{\varphi}(\mathbf{k},z) = \tilde{\varphi}_e(\mathbf{k},0)\frac{\sinh(k(h-z)/\tilde{\gamma})}{\sinh(k\,h/\tilde{\gamma})} + \left(\begin{array}{l}\displaystyle\int_0^z d\xi\,\tilde{P}_3(\mathbf{k},\xi)\frac{\cosh(k\xi/\tilde{\gamma})\sinh(k(h-z)/\tilde{\gamma})}{\varepsilon_0\cdot\sinh(k\,h/\tilde{\gamma})} - \\ \displaystyle\int_z^h d\xi\,\tilde{P}_3(\mathbf{k},\xi)\frac{\sinh(k\,z/\tilde{\gamma})\cosh(k(h-\xi)/\tilde{\gamma})}{\varepsilon_0\cdot\sinh(k\,h/\tilde{\gamma})}\end{array}\right) \tag{A.5a}$$

Corresponding Fourier representation of electric field z-component acquires the form:

$$\tilde{E}_3(\mathbf{k},z) = \left(\begin{array}{l}\tilde{\varphi}_e(\mathbf{k},0)\dfrac{k\cosh(k(h-z)/\tilde{\gamma})}{\tilde{\gamma}\cdot\sinh(k\,h/\tilde{\gamma})} - \dfrac{1}{\varepsilon_0}\tilde{P}_3(\mathbf{k},z) + \\ +\displaystyle\int_0^z d\xi\,\tilde{P}_3(\mathbf{k},\xi)\cosh(k\xi/\tilde{\gamma})\dfrac{k\cosh(k(h-z)/\tilde{\gamma})}{\varepsilon_0\cdot\tilde{\gamma}\cdot\sinh(k\,h/\tilde{\gamma})} + \\ +\displaystyle\int_z^h d\xi\,\tilde{P}_3(\mathbf{k},\xi)\cosh(k(h-\xi)/\tilde{\gamma})\dfrac{k\cosh(k\,z/\tilde{\gamma})}{\varepsilon_0\cdot\tilde{\gamma}\cdot\sinh(k\,h/\tilde{\gamma})}\end{array}\right) \tag{A.5b}$$

In particular case, when the system is transversally uniform one obtains at $k=0$:

$$\varphi(z) = \varphi_e(0)\frac{h-z}{h} + \left(\int_0^z d\xi \frac{P_3(\xi)}{\varepsilon_0} - \frac{z}{h}\int_0^h d\xi \frac{P_3(\xi)}{\varepsilon_0}\right) \tag{A.6a}$$

Corresponding electric field acquires the form:

$$E_3(z) = \frac{\varphi_e(0)}{h} - \left(\frac{P_3(z)}{\varepsilon_0} - \frac{1}{h}\int_0^h d\xi \frac{P_3(\xi)}{\varepsilon_0}\right) \tag{A.6b}$$

The brackets in Eq.(A.6b) is exactly the expression for depolarization field obtained by Kretschmer and Binder.[1]

For the semi-infinite media one can obtain from Eq. (A.5b) in the limit $h \to \infty$ the following expression:

$$\tilde{E}_3(\mathbf{k},z) = \tilde{\varphi}_e(\mathbf{k})\frac{k}{\tilde{\gamma}}\exp(-kz/\tilde{\gamma}) - \frac{1}{\varepsilon_0}\tilde{P}_3(\mathbf{k},z) +$$
$$+ \int_0^{+\infty} d\xi \frac{\tilde{P}_3(\mathbf{k},\xi)}{\varepsilon_0}\frac{k}{\tilde{\gamma}}\left(\frac{\exp(-k|\xi-z|/\tilde{\gamma}) + \exp(-k(\xi+z)/\tilde{\gamma})}{2}\right) \tag{A.7}$$

The first and the second terms in the brackets are related to bulk source and its image in the surface.

### A.2. Elastic problem solution

(a) Compatibility conditions $\mathrm{inc}(i,j,\hat{u}) = e_{ikl}e_{jmn}\left(\partial^2 u_{ln}/\partial x_k \partial x_m\right) = 0$ for the case when all functions depend on $x$ and $z$ only, for the components and along with equations of state $Q_{ij33}P_3^2 + s_{ijkl}\sigma_{kl} = u_{ij}$ lead to and $\sigma_{12,13} - \sigma_{23,11} = 0$ (comma separates the derivatives). Mechanical equilibrium conditions leads to $\sigma_{12,1} + \sigma_{23,3} = 0$, so $\sigma_{12,11} + \sigma_{23,31} = 0$ and $\sigma_{12,13} + \sigma_{23,33} = 0$ after differentiating on $x$ and $z$. From here $\sigma_{12,33} + \sigma_{12,11} = 0$ and $\sigma_{23,33} + \sigma_{23,11} = 0$, which along with boundary conditions $\sigma_{23}(z=0) = 0$ and $\sigma_{23}(z \to \infty) = 0$, $\sigma_{23}(x \to \infty) = 0$, $\sigma_{12}(z \to \infty) = 0$, $\sigma_{12}(x \to \infty) = 0$ lead to $\sigma_{23} = 0$ and $\sigma_{12} = 0$. Then remained equilibrium conditions $\sigma_{11,1} + \sigma_{13,3} = 0$ and $\sigma_{13,1} + \sigma_{33,3} = 0$ can be fulfilled by introducing of elastic potential $\chi(x,z)$ as following:[39]

$$\sigma_{11} = \chi_{,33}(x,z), \quad \sigma_{13} = -\chi_{,13}(x,z), \quad \sigma_{33} = \chi_{,11}(x,z). \tag{A.8}$$

(b) Compatibility conditions for the components $\mathrm{inc}(1,1,\hat{u})$, $\mathrm{inc}(1,3,\hat{u})$ and $\mathrm{inc}(3,3,\hat{u})$ lead to the conditions $u_{22,33} = u_{22,11} = u_{22,13} = 0$, which gives $u_{22} = \mathrm{const}$ owing to the finite strain conditions at infinity. The constant $u_{22}$ should be determined from the corresponding equation of state as $u_{22} = Q_{12} P_S^2$ since stress vanishes and $P_3 \to \pm P_S$ at infinity. Then corresponding equation $u_{22} = Q_{12} P_3^2 + s_{12}(\sigma_{11} + \sigma_{33}) + s_{11}\sigma_{22}$ immediately gives:

$$s_{12}(\sigma_{11} + \sigma_{33}) + s_{11}\sigma_{22} = Q_{12}\left(P_S^2 - P_3^2(x,z)\right). \tag{A.9}$$

(c) Compatibility condition for the component $\mathrm{inc}(2,2,\hat{u}) = u_{11,33} + u_{33,11} - 2u_{13,13} = 0$ after elementary transformations lead to equation for $\chi(x,z)$:

$$\begin{pmatrix}(\chi_{,3333} + \chi_{,1111})(s_{11}^2 - s_{12}^2) + \\ + (2(s_{11} - s_{12})s_{12} + s_{44}s_{11})\chi_{,1133}\end{pmatrix} = -Q_{12}(s_{11} - s_{12})(P_3^2)_{,33} - (Q_{11}s_{11} - Q_{12}s_{12})(P_3^2)_{,11}. \tag{A.10}$$

For elastically isotropic media $2(s_{11} - s_{12})s_{12} + s_{44}s_{11} \to 2(s_{11}^2 - s_{12}^2)$ one obtains well-known biharmonic equation for $\chi(x,z)$. In general case

$$\chi_{,3333} + \chi_{,1111} + 2\gamma_S^2 \chi_{,1133} = \frac{Q_{12}\Delta P_{3,33}^2}{s_{11} + s_{12}} + \frac{Q_{11}s_{11} - Q_{12}s_{12}}{s_{11}^2 - s_{12}^2}\Delta P_{3,11}^2. \tag{A.11}$$

Where the elastic anisotropy factor $\gamma_S^2 = \dfrac{2(s_{11} - s_{12})s_{12} + s_{44}s_{11}}{2(s_{11}^2 - s_{12}^2)}$ and designation $\Delta P_3^2 = P_S^2 - P_3^2(x,z)$ are introduced. In order to solve (A.11) let us use Fourier transformations on coordinate $x$ as $\chi(x,z) = \dfrac{1}{\sqrt{2\pi}} \int\limits_{-\infty}^{\infty} dk_1 \exp(-ik_1 x) \cdot \tilde{\chi}(k_1, z)$ and obtained

$$\frac{d^4\tilde{\chi}}{dz^4} + k_1^4 \tilde{\chi} - 2\gamma_S^2 k_1^2 \frac{d^2\tilde{\chi}}{dz^2} = \frac{Q_{12}}{s_{11} + s_{12}} \frac{d^2 \Delta\tilde{P}_3^2}{dz^2} - \frac{Q_{11}s_{11} - Q_{12}s_{12}}{s_{11}^2 - s_{12}^2} k_1^2 \Delta\tilde{P}_3^2. \tag{A.12}$$

along with the boundary conditions $\tilde{\chi}(k_1, 0) = 0$, $\dfrac{d}{dz}\tilde{\chi}(k_1, 0) = 0$, $\tilde{\chi}(k_1, \infty) = \mathrm{const}$, $\dfrac{d}{dz}\tilde{\chi}(k_1, \infty) = \mathrm{const}$.

Further we consider elastically isotropic case $\gamma_S^2 = 1$ and $P_3^2(x,z) \to P_0^2(x)$ as zero approximation for 1D domain structure. For the case solution of Eq.(A.12) was obtained:

$$\tilde{\chi}(k_1, z) = -\frac{Q_{11}s_{11} - Q_{12}s_{12}}{(s_{11}^2 - s_{12}^2)k_1^2} \Delta \tilde{P}_3^2(k_1, k_2)\left(1 - (1 - |k_1||z|)\exp(-|k_1||z|)\right), \quad \text{(A.13a)}$$

$$\tilde{\sigma}_{33}(k_1, z) = -k_1^2 \tilde{\chi} = \frac{Q_{11}s_{11} - Q_{12}s_{12}}{s_{11}^2 - s_{12}^2} \Delta \tilde{P}_3^2(k_1, k_2)\left(1 - (1 + |k_1||z|)\exp(-|k_1||z|)\right) \quad \text{(A.13b)}$$

$$\tilde{\sigma}_{11}(k_1, z) = \frac{d^2 \tilde{\chi}}{dz^2} = -\frac{Q_{11}s_{11} - Q_{12}s_{12}}{s_{11}^2 - s_{12}^2} \Delta \tilde{P}_3^2(k_1, k_2)\left(1 - |k_1||z|\right)\exp(-|k_1||z|) \quad \text{(A.13c)}$$

$$\tilde{\sigma}_{22}(k_1, z) = \left(\frac{Q_{12}}{s_{11}} - \frac{s_{12}}{s_{11}}\frac{Q_{11}s_{11} - Q_{12}s_{12}}{s_{11}^2 - s_{12}^2} + 2\frac{s_{12}}{s_{11}}\frac{Q_{11}s_{11} - Q_{12}s_{12}}{s_{11}^2 - s_{12}^2}\exp(-|k_1||z|)\right)\Delta \tilde{P}_3^2(k_1, k_2) \quad \text{(A.13d)}$$

Also $\tilde{\sigma}_{13}(k_1, z) \neq 0$ can be easy obtained from (A.8), but it does not contribute into the convolution $Q_{ij33}\tilde{\sigma}_{ij} = Q_{11}\tilde{\sigma}_{33} + Q_{12}(\tilde{\sigma}_{11} + \tilde{\sigma}_{22})$ for cubic symmetry. Far from the surface ($z \to \infty$) original of Eqs.(13) have the form:

$$\sigma_{11}(x, \infty) = 0, \quad \sigma_{22}(x, \infty) = \frac{Q_{12}s_{11} - Q_{11}s_{12}}{s_{11}^2 - s_{12}^2}\left(P_S^2 - P_0^2(x)\right), \quad \text{(A.14a)}$$

$$\sigma_{33}(x, \infty) = \frac{Q_{11}s_{11} - Q_{12}s_{12}}{s_{11}^2 - s_{12}^2}\left(P_S^2 - P_0^2(x)\right). \quad \text{(A.14b)}$$

On the surface ($z=0$) original of Eqs.(13) have the form:

$$\sigma_{11}(x, 0) = -\frac{Q_{11}s_{11} - Q_{12}s_{12}}{s_{11}^2 - s_{12}^2}\left(P_S^2 - P_0^2(x)\right), \quad \sigma_{33}(x, 0) = 0, \quad \text{(A.15a)}$$

$$\sigma_{22}(x, 0) = \left(\frac{Q_{12}}{s_{11}} + \frac{s_{12}}{s_{11}}\frac{Q_{11}s_{11} - Q_{12}s_{12}}{s_{11}^2 - s_{12}^2}\right)\left(P_S^2 - P_0^2(x)\right), \quad \text{(A.15b)}$$

Finally convolution $Q_{ij33}\sigma_{ij} = Q_{11}\sigma_{33} + Q_{12}(\sigma_{11} + \sigma_{22})$ has the form:

$$Q_{ij33}\sigma_{ij}(x, \infty) = \frac{(Q_{11}^2 + Q_{12}^2)s_{11} - 2Q_{12}Q_{11}s_{12}}{s_{11}^2 - s_{12}^2}\left(P_S^2 - P_0^2(x)\right) \quad \text{(A.16a)}$$

$$Q_{ij33}\sigma_{ij}(x, 0) = Q_{12}\frac{Q_{12}(s_{11} + 2s_{12}) - Q_{11}s_{11}}{s_{11}(s_{11} + s_{12})}\left(P_S^2 - P_0^2(x)\right) \quad \text{(A.16b)}$$

**Appendix B. Solution of linearized equation.**

The free energy functional (5) acquires the following form for Fourier images:

$$G \approx \int_{-\infty}^{\infty} dk_1 \int_{-\infty}^{\infty} dk_2 \left( \begin{array}{l} \int_0^h dz \left[ \begin{array}{l} \dfrac{\alpha}{2}\left|\tilde{P}_3(\mathbf{k},z)\right|^2 + \dfrac{\beta}{4}\left|\tilde{P}_3^2(\mathbf{k},z)\right|^2 + \dfrac{\delta}{6}\left|\tilde{P}_3^3(\mathbf{k},z)\right|^2 + \\ + \dfrac{\zeta}{2}\left|\dfrac{\partial \tilde{P}_3(\mathbf{k},z)}{\partial z}\right|^2 + \dfrac{\eta}{2}k^2\left|\tilde{P}_3(\mathbf{k},z)\right|^2 - \tilde{P}_3(\mathbf{k},z)\tilde{E}_3^{*e}(\mathbf{k},z) - \\ -\dfrac{1}{2}\tilde{P}_3(\mathbf{k},z)\tilde{E}_3^{*d}(\mathbf{k},z) - Q_{ij33}\tilde{\sigma}_{ij}^*(\mathbf{k},z)\tilde{P}_3^2(\mathbf{k},z) - \dfrac{s_{ijkl}}{2}\tilde{\sigma}_{ij}(\mathbf{k},z)\tilde{\sigma}_{kl}^*(\mathbf{k},z) \end{array} \right] \\ + \dfrac{\zeta}{2\lambda}\left(\left|\tilde{P}_3(\mathbf{k},0)\right|^2 + \left|\tilde{P}_3(\mathbf{k},h)\right|^2\right) \end{array} \right) \quad (B.1)$$

Here we used Parseval theorem, identity $\tilde{E}_3^{d,e}(-\mathbf{k},z) = \tilde{E}_3^{*d,e}(\mathbf{k},z)$ ("e" is external, "d" is depolarization field). Note, that Bogolubov approximation for Fourier images of the terms $P_3^4$ and $P_3^6$ leads to $\left|\tilde{P}_3^2(\mathbf{k},z)\right|^2 \to \left|\tilde{P}_3(\mathbf{k},z)\right|^4$ and $\left|\tilde{P}_3^3(\mathbf{k},z)\right|^2 \to \left|\tilde{P}_3(\mathbf{k},z)\right|^6$. Variation of Eq.(B.1) leads to:

$$\alpha_S \tilde{P}_3(\mathbf{k},z) + \beta_S \tilde{P}_3^3(\mathbf{k},z) + \delta \tilde{P}_3^5(\mathbf{k},z) - \left(\zeta\dfrac{\partial^2}{\partial z^2} - \eta k^2\right)\tilde{P}_3(\mathbf{k},z) = \tilde{E}_3(\mathbf{k},z) \quad (B.2a)$$

along with the boundary conditions

$$\left(\tilde{P}_3(\mathbf{k},z) - \lambda\dfrac{\partial \tilde{P}_3(\mathbf{k},z)}{\partial z}\right)\bigg|_{z=0} = 0, \quad \left(\tilde{P}_3(\mathbf{k},z) + \lambda\dfrac{\partial \tilde{P}_3(\mathbf{k},z)}{\partial z}\right)\bigg|_{z=h} = 0. \quad (B.2b)$$

Let us apply operator $\dfrac{d^2}{dz^2} - \dfrac{k^2}{\tilde{\gamma}^2}$ to the field $\tilde{E}_3(\mathbf{k},z)$. After simple but cumbersome transformations one obtains that $\left(\dfrac{d^2}{dz^2} - \dfrac{k^2}{\tilde{\gamma}^2}\right)\tilde{E}_3(\mathbf{k},z) = -\dfrac{1}{\varepsilon_0}\dfrac{d^2 \tilde{P}_3(\mathbf{k},z)}{dz^2}$ as anticipated directly from (2). Then applying operator $\dfrac{d^2}{dz^2} - \dfrac{k^2}{\tilde{\gamma}^2}$ to linearized Eq. (B.2a), we obtained

$$\left(\dfrac{d^2}{dz^2} - \dfrac{k^2}{\tilde{\gamma}^2}\right)\left[\left(\alpha_S + 3\beta_S P_S^2 + 5\delta P_S^4\right) - \left(\zeta\dfrac{d^2}{dz^2} - \eta k^2\right)\right]p(\mathbf{k},z) = -\dfrac{1}{\varepsilon_0}\dfrac{d^2 p(\mathbf{k},z)}{dz^2} \quad (B.3)$$

Where $\tilde{P}_3(\mathbf{k},z) \approx P_0(\mathbf{k}) + p(\mathbf{k},z)$. Eq (B.3) along with the boundary conditions (7) can be rewritten as

$$\left(\dfrac{d^2}{dz^2} - \dfrac{k^2}{\gamma^2}\varepsilon_{33}\right)\left[1 - \left(L_\perp^2\dfrac{d^2}{dz^2} - L_z^2 k^2\right)\right]p(\mathbf{k},z) = -(\varepsilon_{33} - 1)\dfrac{d^2 p(\mathbf{k},z)}{dz^2} \quad (B.4)$$

$$\left(p(\mathbf{k},z)-\lambda\frac{\partial p(\mathbf{k},z)}{\partial z}\right)\bigg|_{z=0}=-P_0(\mathbf{k}), \quad \left(p(\mathbf{k},z)+\lambda\frac{\partial p(\mathbf{k},z)}{\partial z}\right)\bigg|_{z=h}=-P_0(\mathbf{k}). \quad (B.5)$$

Here correlation lengths are introduced as

$$L_z=\sqrt{\frac{\zeta}{\alpha_S+3\beta_S P_S^2+5\delta P_S^4}}, \quad L_\perp=\sqrt{\frac{\eta}{\alpha_S+3\beta_S P_S^2+5\delta P_S^4}} \quad (B.6)$$

Looking for the solution of Eq. (B.4) in the form $p(\mathbf{k},z) \sim \exp(s\,z)$, one can find characteristic equation for $s$ in the form:

$$\left(s^2-\frac{k^2}{\gamma^2}\varepsilon_{33}\right)\!\left(1-\left(L_z^2 s^2-L_\perp^2 k^2\right)\right)=-(\varepsilon_{33}-1)s^2 \quad (B.7)$$

Here $\gamma=\sqrt{\varepsilon_{33}/\varepsilon_{11}}$ and $\varepsilon_{33}=1+\dfrac{1}{\varepsilon_0\left(\alpha_S+3\beta_S P_S^2+5\delta P_S^4\right)}$. The roots of this biquadratic equation are

$$s_{1,2}^2=\frac{\varepsilon_{33}(\gamma^2+k^2 L_z^2)+\gamma^2 L_\perp^2 k^2 \pm\sqrt{\left(\varepsilon_{33}(\gamma^2+k^2 L_z^2)+\gamma^2 L_\perp^2 k^2\right)^2-4\varepsilon_{33}\gamma^2 L_z^2 k^2(1+k^2 L_\perp^2)}}{2L_z^2\gamma^2} \quad (B.8)$$

It is seen that for any real values of $k$ values of $s_{1,2}$ are real. So that the general solution of Eq. (B.5) acquires the form

$$p(\mathbf{k},z)=A_1\cosh(s_1 z)+B_1\cosh(s_1(h-z))+A_2\cosh(s_2 z)+B_2\cosh(s_2(h-z)) \quad (B.9)$$

After substitution into (7a) one obtains:

$$\begin{aligned}0=&(\varepsilon_{33}-1)\cdot\varepsilon_0\tilde{\varphi}_e(\mathbf{k})\frac{k\cosh(k(h-z)/\tilde{\gamma})}{\tilde{\gamma}\cdot\sinh(k\,h/\tilde{\gamma})}-\\ &-(\varepsilon_{33}-1)\frac{k\,s_1\tilde{\gamma}}{k^2-s_1^2\tilde{\gamma}^2}\frac{\sinh(s_1 h)}{\sinh(k\,h/\tilde{\gamma})}\left(A_1\cosh\!\left(\frac{k}{\tilde{\gamma}}z\right)+B_1\cosh\!\left(\frac{k}{\tilde{\gamma}}(h-z)\right)\right)-\\ &-(\varepsilon_{33}-1)\frac{k\,s_2\tilde{\gamma}}{k^2-s_2^2\tilde{\gamma}^2}\frac{\sinh(s_2 h)}{\sinh(k\,h/\tilde{\gamma})}\left(A_2\cosh\!\left(\frac{k}{\tilde{\gamma}}z\right)+B_2\cosh\!\left(\frac{k}{\tilde{\gamma}}(h-z)\right)\right)\end{aligned} \quad (B.10)$$

Eq.(B.10) along with the boundary conditions (B.5b) leads to the system of equations for constants $A_i$ and $B_i$:

$$\frac{k\,s_1\tilde{\gamma}}{k^2-s_1^2\tilde{\gamma}^2}\sinh(s_1 h)B_1+\frac{k\,s_2\tilde{\gamma}}{k^2-s_2^2\tilde{\gamma}^2}\sinh(s_2 h)B_2=\varepsilon_0\tilde{\varphi}_e(\mathbf{k})\frac{k}{\tilde{\gamma}} \quad (B.11a)$$

$$\frac{k\,s_1\tilde{\gamma}}{k^2-s_1^2\tilde{\gamma}^2}\sinh(s_1 h)A_1+\frac{k\,s_2\tilde{\gamma}}{k^2-s_2^2\tilde{\gamma}^2}\sinh(s_2 h)A_2=0 \quad (B.11b)$$

$$\begin{pmatrix} A_1 + B_1 \cosh(s_1 h) + A_2 + B_2 \cosh(s_2 h) + \\ + \lambda B_1 s_1 \sinh(s_1 h) + \lambda B_2 s_2 \sinh(s_2 h) \end{pmatrix} = -P_0(\mathbf{k}) \qquad \text{(B.11c)}$$

$$\begin{pmatrix} A_1 \cosh(s_1 h) + B_1 + A_2 \cosh(s_2 h) + B_2 + \\ + \lambda A_1 s_1 \sinh(s_1 h) + \lambda A_2 s_2 \sinh(s_2 h) \end{pmatrix} = -P_0(\mathbf{k}) \qquad \text{(B.11d)}$$

In particular case $\widetilde{\varphi}_e(\mathbf{k}) = 0$ its solution has form:

$$A_1(s_1, s_2) \equiv B_1(s_1, s_2) = \frac{P_0(\mathbf{k}) \cdot \sinh(s_2 h/2) M(s_2)}{\cosh(s_1 h/2) Det_I(s_1, s_2, h)}, \qquad \text{(B.12a)}$$

$$A_2(s_1, s_2) \equiv B_2(s_1, s_2) = -\frac{P_0(\mathbf{k}) \sinh(s_1 h/2) M(s_1)}{\cosh(s_2 h/2) Det_I(s_1, s_2, h)}, \qquad \text{(B.12b)}$$

$$Det_I(s, q, h) = 2 \begin{pmatrix} M(s) \cosh\left(\frac{q h}{2}\right) \sinh\left(\frac{s h}{2}\right) - M(q) \cosh\left(\frac{s h}{2}\right) \sinh\left(\frac{q h}{2}\right) + \\ + \lambda (q M(s) - s M(q)) \sinh\left(\frac{s h}{2}\right) \sinh\left(\frac{q h}{2}\right) \end{pmatrix} \qquad \text{(B.12c)}$$

Where $M(s) = \dfrac{s}{\varepsilon_{11} k^2 - s^2}$. It is clear that $A_1(s_1, s_2) = -A_2(s_2, s_1)$.

At a given extrapolation length $\lambda$, linearized solution of the system diverges at several $k$ values determined from the condition $Det_I(s_1(k), s_2(k), \lambda, h) = 0$. Corresponding solution $\lambda_{cr}(k)$ or $k_{cr}(\lambda)$ indicates the instability point of bulk domain structure $P_0(x)$ with period $2\pi/k_{cr}(\lambda)$ induced by the surface influence. Dependence $\lambda_{cr}(k)$ is shown in Fig.B1 for typical ferroelectrics material parameters and different thickness $h$.

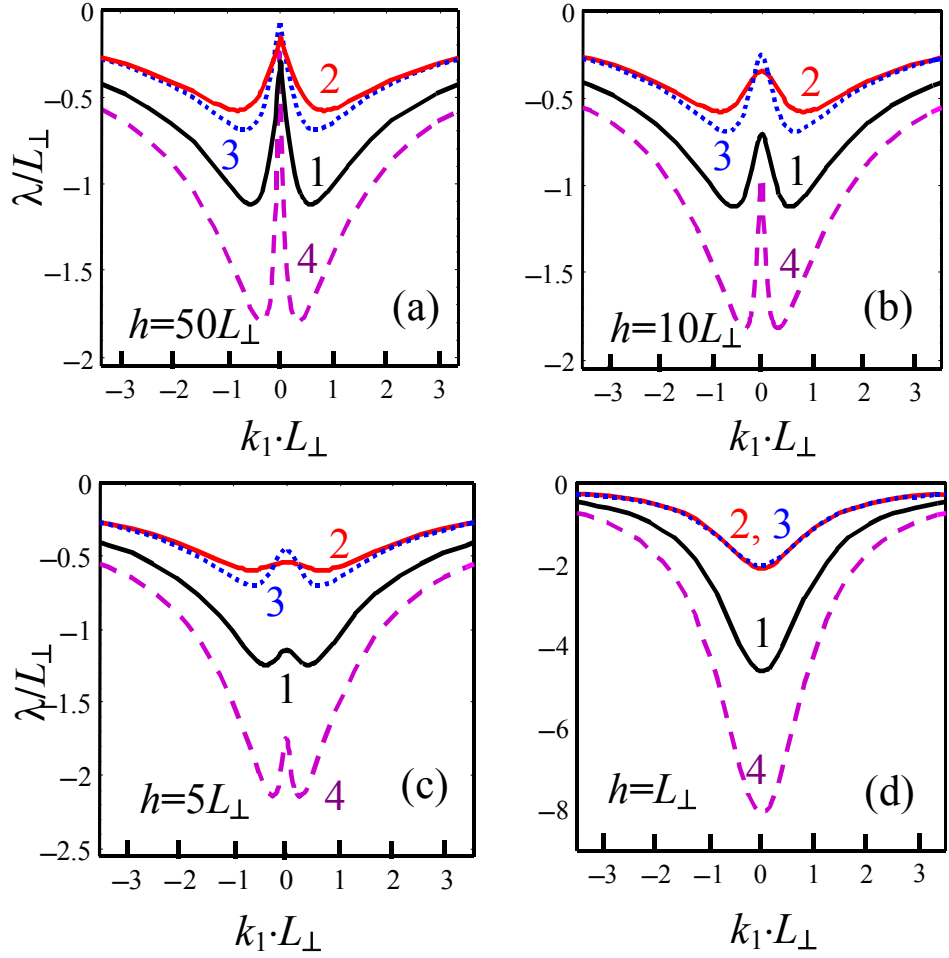

**Figure B1.** (Color online) Dependence $\lambda_{cr}(k_1)$ calculated from Eq.(B.12c) in LiNbO$_3$ at $L_z/L_\perp=1.5$ (curves 1), LiTaO$_3$ at $L_z/L_\perp=1$ (curves 2), PbZr$_{0.5}$Ti$_{0.5}$O$_3$ at $L_z/L_\perp=1$ (curves 3) and BaTiO$_3$ at $R_L/L_\perp=2$ (curves 4) for different film thickness $h/L_\perp=50, 10, 5, 1$ (parts a, b, c, d).

It is clear that zero determinant $Det_I$ given by Eq.(B.12c) is possible only at negative $\lambda$ values, at that two maximums $\lambda_{cr}(\pm k_1)$ exist in semi-infinite sample and thick films as shown in Figs.B1a-c; they split into the single maxima $\lambda_{cr}(0)$ with film thickness decrease as shown in Fig. B1d. Note, that thickness-induced paraelectric phase transition at $h<h_{cr}$ takes place only at $\lambda \geq 0$.

The considered spontaneous stripe domain splitting near the film surface appeared at negative $\lambda$ values could not be treated in terms of linearized approach (13), however the condition $Det_I(s_1(k), s_2(k), \lambda, h) = 0$ determine the most probable structure period at a given $\lambda$. Then, in order to determine the polarization amplitude direct variational method should be used.

Below we consider the range of extrapolation length values where the bulk domain structure $P_0(x)$ is stable and so one may suspect $P_V \approx 1$ to be a good approximation (e.g $\lambda > 0$ and $\lambda < -2$ for thickness $h > 10 L_\perp$).

For particular case $\mathbf{k} \to 0$ (transversally homogeneous film) one can obtain characteristic increments as $s_{10}^2 = \frac{\varepsilon_{11} k^2}{\varepsilon_{33}} \to 0$, $s_{20}^2 = \frac{\varepsilon_{33}}{L_z^2}\left(1 + \frac{k^2}{\varepsilon_{33}}\left(L_\perp^2 + \frac{\varepsilon_{11}}{\varepsilon_{33}}(\varepsilon_{33}-1)L_z^2\right)\right) \approx \frac{1}{\varepsilon_0 \zeta}$ and determinant $Det_I(s_{10}, s_{20}, h) = 2\left(\frac{1}{s_{20}}\sinh\left(\frac{s_{20} h}{2}\right) + \frac{h}{2(\varepsilon_{33}-1)}\left(\cosh\left(\frac{s_{20} h}{2}\right) + \lambda s_{20} \sinh\left(\frac{s_{20} h}{2}\right)\right)\right)$. Under the typical condition $s_{20} h \gg 1$, the solution is

$$\tilde{P}_3(\mathbf{k}=0, z) \approx P_S\left(1 + \frac{2(\varepsilon_{33}-1)\tanh(s_{20} h/2)}{s_{20} h(1 + \lambda s_{20})\tanh(s_{20} h/2)}\right)^{-1}\left(1 - \frac{\cosh(s_{20}(h/2-z))}{\cosh(s_{20} h/2) + \lambda s_{20} \sinh(s_{20} h/2)}\right),$$

(B.13)

For particular case $h \to \infty$ one obtains determinant $Det_I(s, q, h \to \infty) \to \frac{1}{2}(M(s) - M(q) + \lambda(q M(s) - s M(q)))\exp\left(\frac{(q+s)h}{2}\right)$, coefficient $A(s, q) \to \frac{\tilde{P}_0(\mathbf{k}) \cdot 2\exp(-s h)M(q)}{M(s) - M(q) + \lambda(q M(s) - s M(q))}$, polarization Fourier image $\tilde{P}_3(\mathbf{k}, z) = \tilde{P}_0(\mathbf{k})\left(1 + \frac{P_V(\exp(-s_1 z)M(s_2) - \exp(-s_2 z)M(s_1))}{M(s_1) - M(s_2) + \lambda(s_2 M(s_1) - s_1 M(s_2))}\right)$ explicit form is:

$$\tilde{P}_3(\mathbf{k}, 0) = \tilde{P}_0(\mathbf{k})\left(1 + P_V \frac{\exp(-s_1 z)(\varepsilon_{11} k^2 - s_1^2)s_2 - \exp(-s_2 z)(\varepsilon_{11} k^2 - s_2^2)s_1}{(s_1 - s_2)(\varepsilon_{11} k^2 + s_1 s_2 + \lambda s_1 s_2(s_1 + s_2))}\right).$$ (B.14)

At the surface ($z=0$) polarization $\tilde{P}_3(\mathbf{k}, 0) = \tilde{P}_0(\mathbf{k})\left(1 - \frac{(\varepsilon_{11} k^2 + s_1 s_2)P_V}{\varepsilon_{11} k^2 + s_1 s_2 + \lambda s_1 s_2(s_1 + s_2)}\right)$, where

$s_1 s_2 \equiv \sqrt{\frac{\varepsilon_{11} k^2 (1 + R_T^2 k^2)}{R_L^2}}$ and $s_1 + s_2 \equiv \sqrt{\frac{\varepsilon_{33} + k^2(R_T^2 + \varepsilon_{11} R_L^2)}{R_L^2} + 2\sqrt{\frac{\varepsilon_{11} k^2(1 + R_T^2 k^2)}{R_L^2}}}$ In accordance with Eq.(B.7).

Approximate analytical results can be derived for a single domain wall profile in the second order ferroelectrics, since $\tilde{P}_0(\mathbf{k}) = \dfrac{i\pi R_T \, \text{Delta}(k_2) \cdot P_S}{\sinh(\pi k_1 R_T)}$, where Delta(**k**) is Dirac-delta function.

For odd functions $f_0(x) = \int\limits_{-\infty}^{\infty} \tilde{f}_0(k_1)\exp(ik_1 x)dk_1$, so

$$f(x,z) = \int\limits_{-\infty}^{\infty} \tilde{f}_0(k_1)\frac{|k_1|\exp(ik_1 x - |k_1|zC)}{1 + q|k_1| + Q|k_1|^2 + \ldots} dk_1 \rightarrow \int\limits_{-\infty}^{\infty} \tilde{f}_0(k_1)|k_1|\exp(ik_1 x - (zC+q)|k_1|) dk_1 =$$

$$= -\frac{d}{Cdz}\int\limits_{-\infty}^{\infty} \tilde{f}_0(k_1)\exp(ik_1 x - |k_1|(zC+q)) dk_1 =$$

$$= \int\limits_{-\infty}^{\infty} dy f_0(x-y)\frac{-d}{Cdz}\frac{2(zC+q)}{y^2 + (zC+q)^2} = \int\limits_{-\infty}^{\infty} dy f_0(x-y)\frac{2((zC+q)^2 - y^2)}{(y^2 + (zC+q)^2)^2} =$$

$$= \int\limits_{-\infty}^{\infty} dy f_0(x-y)\left(\frac{-2}{y^2 + (zC+q)^2} + \frac{4(zC+q)^2}{(y^2 + (zC+q)^2)^2}\right)$$

(B.15)

**Appendix C. Phase field modeling**

Below we study numerically the effect of finite extrapolation length on periodic *c*-domain structure near the surfaces of a thin film by using phase field method. The spontaneous polarization, $\mathbf{P}=(P_1, P_2, P_3)$, is taken as the order parameter. For the considered uniaxial ferroelectrics LiTaO$_3$ and LiNbO$_3$, $P_1=P_2=0$ are assumed. The spatial-temporal evolution for $P_3$ is calculated from the Landau-Khalatnikov equation

$$\frac{\partial P_3(\mathbf{r},t)}{\partial t} = -\Gamma \frac{\delta G}{\delta P_3(\mathbf{r},t)},$$
(C.1)

where $\Gamma$ is the kinetic coefficient, related to the domain wall mobility, radius-vector $\mathbf{r}=(x,y,z)$, *G* (or *F* depending on the mechanical boundary conditions) is the free energy of the system given by Eq.(5). Variational derivative $\delta G/\delta P_3(\mathbf{r},t)$ represents the thermodynamic driving force for the spatial and temporal evolution of the simulated system.

Corresponding boundary conditions are $\left(P_3 - \lambda_1 \dfrac{\partial P_3}{\partial z}\right)\bigg|_{z=0} = 0, \left(P_3 + \lambda_2 \dfrac{\partial P_3}{\partial z}\right)\bigg|_{z=h} = 0$.

The free energy bulk density *g* includes polarization (or Landau) energy, domain wall (or correlation) energy and electrostatic energy. For 180°-domain wall in LiTaO$_3$ or LiNbO$_3$ the elastic energy contribution appeared relatively small allowing for small striction coefficients (see also Table 1). So the free energy density is written as

$$g = f_{Lan}(P_3) + f_{grad}(P_{3,j}) + f_{elec}(E_3, P_3), \quad (C.2)$$

where $f_{Lan} = \frac{\alpha}{2}P_3^2 + \frac{\beta}{4}P_3^4$. The expansion coefficients in SI units are $\alpha$=-1.256×10$^9$, $\beta$=5.043×10$^9$ for LiTaO$_3$ and $\alpha$=-2.012×10$^9$, $\beta$=3.608×10$^9$ for LiNbO$_3$, respectively.

The correlation energy density is $f_{grad} = \frac{1}{2}\zeta\left(\frac{\partial P_3}{\partial z}\right)^2 + \frac{1}{2}\eta\left[\left(\frac{\partial P_3}{\partial x}\right)^2 + \left(\frac{\partial P_3}{\partial y}\right)^2\right]$, where $\eta$ and $\zeta$ are the gradient energy coefficients. In the simulations, we take $\eta = -\frac{1}{2}\eta^*\alpha L^2$ and $\zeta = -\frac{1}{2}\zeta^*\alpha H^2$, where $\eta^*$ and $\zeta^*$ are dimensionless parameters, $H$ and $L$ represent the real simulation cell size of $2\pi L \times 2H$ in a 2D model, and $\zeta^* = \frac{4L_z^2}{H^2}$, $\eta^* = \frac{4L_\perp^2}{L^2}$.

The electrostatic energy density, which can be expressed as $f_{elec} = -\left(E_3^0 + \frac{1}{2}E_3^d\right)P_3$, where $E_3^d$ is the component of the depolarization electric field. Without any applied electric field $E_3^0$, depolarization field is induced only by the inhomogeneous spontaneous polarizations allowing for screening charges on the electrodes. Depolarization field potential φ satisfy electrostatic equilibrium equation (2), namely $\frac{\partial^2 \varphi}{\partial z^2} + \varepsilon_{11}\left(\frac{\partial^2 \varphi}{\partial x^2} + \frac{\partial^2 \varphi}{\partial y^2}\right) = \frac{1}{\varepsilon_0}\frac{\partial P_3}{\partial z}$ (where $\varepsilon_0 = 8.85 \times 10^{-12}$ Fm$^{-1}$ and $\varepsilon_{11}$=54 for LiTaO$_3$ and $\varepsilon_{11}$=85 for LiNbO$_3$) and short-circuit boundary condition $\varphi|_{z=0} = \varphi|_{z=h} = 0$.

Eq. (C.1) was solved by using a mixed Chebyshev-collocation Fourier-Galerkin method.[55, 56] The simulations started from a 180° periodic domain structure with sharp interface and uniform polarization at each domain. We assumed that electric equilibrium is established instantaneously for a given polarization distribution. The polarization profiles of Fig. 7 are the stable profiles that existed at the end of each simulation at times *t* much longer that Khalatnikov relaxation time.